\documentclass[pre,aps,floatfix,superscriptaddress,two column]{revtex4-2}
\usepackage{graphicx}  
\usepackage{amsmath}   
\usepackage{amssymb}   
\usepackage{bm}        
\usepackage{hyperref}  
\usepackage{physics}
\usepackage{graphicx}
\hypersetup{colorlinks=true, urlcolor=blue, citecolor=blue}
\newcommand\numberthis{\addtocounter{equation}{1}\tag{\theequation}}

\begin{document}

\title{
Analytical Theory of Chiral Active Particle Transport in a Fluctuating Density Field}
\author{Jayam Joshi}
\email{jayam@uchicago.edu}
\affiliation{Department of Physics, The University of Chicago, Chicago, IL 60637.}
\author{Abhra Puitandy}
\email{abhrapuitandy.rs.phy22@itbhu.ac.in}
\affiliation{Department of Physics,Indian Institute of Technology(BHU), Varanasi-221005,India}
\author{Shradha Mishra}
\email{smishra.phy@iitbhu.ac.in}
\affiliation{Department of Physics,Indian Institute of Technology(BHU), Varanasi-221005,India}


\begin{abstract}
We develop a closed-form analytical theory for the transport of a chiral active Brownian particle  in three dimensions, moving through a fluctuating local density field that models steric and dynamical interactions in a dense active medium. The density field is modeled as an Ornstein--Uhlenbeck process with finite correlation time $\tau$ and fluctuation strength $\sigma_\rho^2$, capturing both spatial fluctuations and temporal memory. Within this framework, we derive exact expressions for the mean-squared displacement  and time-dependent diffusivity, revealing how chirality and density coupling jointly renormalise orientational persistence and generate nontrivial dynamical crossovers. The theory predicts: (\textit{i}) anomalously high initial diffusivity for particles starting in locally denser regions, arising from a transient active drift driven by local swim-pressure gradients; (\textit{ii}) a finite crossover time $t_c$ for homogenising density inhomogeneities, with a transient dependence of the dynamics on the initial local density environment which arises from the non-equilibrium evolution of density fluctuations and does not persist when averaging over stationary initial conditions ($\rho_0 = \rho_\infty$) ; (\textit{iii}) a non-monotonic $t_c(\Omega)$ with a global minimum at intermediate chirality, and a three-regime suppression of long-time diffusivity $D_\infty(\Omega)$, consistent with micro-clustered phases observed in simulations; and (\textit{iv}) a resonance-like peak in the early-time oscillatory strength of the mean-squared displacement at an optimal chirality $\Omega^*$, set by the interplay of orientational diffusion, density-field decorrelation, and imposed rotation. The framework captures the qualitative dependence of $D_\infty$ on $Pe$ and $\Omega$, {where Pe denotes the Péclet number}, while uncovering chirality-dependent transport features in active matter, offering a qualitative perspective for understanding transport in biological systems and suggesting possible directions for designing active materials.
\end{abstract}

\maketitle

\section{Introduction}
Active matter systems, comprising self-driven units that consume energy at the microscopic scale, exhibit a rich spectrum of non-equilibrium phenomena including collective motion, clustering, and anomalous transport \cite{Marchetti2013, Elgeti2015, Bechinger2016}. A foundational model for such systems is the active Brownian particle, in which persistent self-propulsion is combined with rotational diffusion \cite{Howse2007, Romanczuk2012}. The theoretical framework of active Brownian particles has yielded deep insights into motility-induced phase separation \cite{Fodor2016, Cates2015}, swim pressure \cite{Takatori2014}, and steady-state currents \cite{solon2015pressure}, particularly in spatially homogeneous settings.

However, real-world active systems often operate in crowded and heterogeneous environments, where local density fluctuations and steric interactions fundamentally reshape particle trajectories. In such contexts, especially in dense suspensions, effective single-particle dynamics are influenced by dynamically evolving local microstructure. These effects become even more intricate in the case of chiral active Brownian particles, where a constant intrinsic torque induces helical propulsion paths \cite{Kummel2013, vanTeeffelen2008, Liebchen2017}. Chirality introduces rotational memory into the dynamics, breaking time-reversal symmetry and generating coupling between translation and rotation over multiple time scales.

Despite increasing interest in chirality-induced effects, theoretical understanding of chiral active Brownian particle transport in three dimensions remains limited. Prior work has focused largely on either dilute regimes or numerical simulation, with few analytical results accounting for crowding and fluctuation effects. In this work, we develop a minimal yet predictive analytical theory that captures the effective dynamics of a single chiral Active Brownian Particle
in a dense medium. We model steric interactions through a stochastic local density field governed by an Ornstein-Uhlenbeck process, incorporating both spatial heterogeneity and finite memory.

This framework allows us to derive exact expressions for the mean-squared displacement  of the tagged particle, revealing how chirality and density coupling modify orientational persistence, enhance transient drift, and influence transient to steady crossover dynamics. The resulting expressions connect microscopic parameters—such as chirality strength, activity, and density statistics—to observable transport coefficients. Our results offer theoretical insight into experimentally accessible systems, including magnetically rotated colloids \cite{tierno2008magnetically}, optically driven microswimmers \cite{maggi2017memory}, and motile bacteria in porous environments \cite{bhattacharjee2019bacterial}, and lay a foundation for generalizing transport theories in structured, active media.

\section{Model}
We first consider a system of $N$ chiral active Brownian particles in three dimensions (3D), where each particle $i$ is characterized by its position $\bm{r}_i(t)$ and orientation $\bm{\hat{n}}_i(t)$. The orientation vector evolves stochastically due to thermal fluctuations and chirality is introduced through an external torque acting on the active Brownian particles (ABPs). The particles interact via soft repulsion. The dynamics of the $i$-th particle are governed by the overdamped Langevin equations of motion:

\begin{equation}
    \partial_{t}{\bm{r}}_i = v_0 \bm{\hat{n}}_i  + \mu \sum_{j \neq i} \bm{F}_{ij} + \sqrt{2 D}\bm{\xi_i}(t)
    \label{eq:position_dynamics}
    \tag{a}
\end{equation}

\begin{equation}
    \partial_{t}{\bm{\hat{n}}}_i = \bm{\hat{n}}_i \times \omega_0 \bm{\hat{z}} + \sqrt{2D_r} \bm{\eta}_i(t) \times \bm{\hat{n}}_i
    \label{eq:orientation_dynamics}
    \tag{b}
\end{equation}
where $v_0$ is the self-propulsion speed, $\mu$ the mobility. The interparticle steric force $\bm{F}_{ij}$ is due to a short-ranged repulsive potential such as the Weeks-Chandler-Andersen (WCA) \cite{weeks1971role} with a cut-off distance.  $\omega_0$ is the chiral frequency inducing precession about the $z$-axis. The position of the particle is subject to translational diffusion with $D$ as the diffusion constant. $\bm{\xi}_i(t)$ representing Gaussian white noise with zero mean and correlation $\langle \xi_{i\alpha}(t) \xi_{j\beta}(t') \rangle = \delta_{ij}\delta_{\alpha\beta}\delta(t-t')$. Similarly, the orientation vector undergoes angular diffusion with $D_r$ as the diffusion constant. $\bm{\eta}_i(t)$ representing Gaussian white noise with zero mean and correlation $\langle \eta_{i\alpha}(t) \eta_{j\beta}(t') \rangle = \delta_{ij}\delta_{\alpha\beta}\delta(t-t')$. The chiral term $\bm{\hat{n}}_i \times \omega_0\bm{\hat{z}}$ in Eq.~(\ref{eq:orientation_dynamics}) breaks time-reversal symmetry, leading to helical trajectories with curvature $\kappa = \omega/v_0$. \\

The transport properties of a chiral ABP (cABP) navigating in a dense suspension of other such particles are affected by steric interactions with neighboring particles.  `{As it is reported in previous study of \cite{stenhammar2014phase}, that in both two and three dimensions the effective speed of the ABP can be written as a functional of local density, which led to the local trapping of the particles and finally the phase separation. This led us to follow the same approach as we did in our previous study \cite{Joshi2025}.  We replace the effect of interactions through a density-dependent self-propulsion speed $v_{\rho}(t)$, where the time-varying field $\rho(t)$ encapsulates effective local density fluctuations around a particle navigating in the environment.} Therefore, we replace the interaction forces $F_{ij}$ in eqn. (\ref{eq:position_dynamics}) and formulate the effective-single cABP Langevin equations of motion:
\begin{equation}
    \partial_{t}{\bm{r}} = v_{\rho}(t) \bm{\hat{n}} + \sqrt{2 D}\bm{\xi}(t) 
    \label{eqn:pos}
\end{equation}
\begin{equation}
     \partial_{t}{\bm{\hat{n}}} = \bm{\hat{n}} \times \omega_0 \bm{\hat{z}} + \sqrt{2D_r} \bm{\eta}(t) \times \bm{\hat{n}}
     \label{eqn:angle}
\end{equation}
where we define self-propulsion speed $v_{\rho}(t)$ as a piece-wise defined function such that: For dilute systems where global packing density $\phi < \phi_c$: 
\begin{equation}
v_{\rho}(t) = v_0 \big(1 - \lambda \rho (t)\big); \lambda>0
\end{equation}
and for dense systems where $\phi > \phi_c$: 
\begin{equation}
    v_{\rho}(t) = v_0 \big(1 - \lambda_1 \rho (t) + \lambda_2 \rho(t)^2\big); \lambda_1, \lambda_2>0
\end{equation}  
This modeling is developed by the observed dependence of effective particle speed on global packing density in \cite{Fily2012}, where speed decreases approximately linearly with density in the homogeneous phase, but exhibits a nonlinear decline beyond a critical density $\phi_c$, characteristic of motility-induced phase separation (MIPS).  Here $\lambda_1$ and $\lambda_2$ are phenomenological coefficients encoding the linear and nonlinear suppression of propulsion speed by local crowding, whose values can be independently extracted from the speed–density relation of dense active suspensions.

More specifically, in the homogeneous phase ($\phi<\phi_c$), particles slow down only by independent, pairwise collisions, so the speed reduction scales linearly, $v(\rho)\approx v_0\bigl(1-\lambda \rho\bigr)$, with each incremental neighbor contributing the same marginal slowdown such that curvature $v^{\prime \prime}(\rho)=0$. As $\phi$ crosses the MIPS threshold, clusters start nucleating  via two‑particle blockade events whose rate grows as $\rho^2$.
The particles become trapped in emergent clusters, each additional neighbor only slightly increases confinement, so the marginal slowdown per particle diminishes leading to a sudden change from zero to positive curvature $v ^{\prime \prime}(\rho) > 0$.  \\
Thus, the piecewise construction of $v_\rho$ captures this transition from homogeneous to phase-separated behavior. The statistical properties of the local density stochastic variable  $\rho(t)$ are analyzed in  \cite{Joshi2025}. From the microscopic dynamics in the system of sterically interacting ABPs on a 2-dimensional substrate, it was found that the stochastic local density $\rho(t)$ exhibit finite memory and approximately exponential temporal correlations which motivated us to model the local density $\rho(t)$ as an Ornstein–Uhlenbeck (OU) process. We emphasize that this is a minimal {\it phenomenological} ansatz intended to capture the effective fluctuating environment experienced by a tagged particle, rather than a quantitatively established description for the specific 3D chiral system. $\rho(t)$ fluctuates around a mean value that relaxes toward a constant value $\rho_\infty$ in the non-equilibrium steady state. In our previous study~\cite{Joshi2025}, we modeled the local density $\rho(t)$ as a stationary Ornstein–Uhlenbeck (OU) process under the assumption of an established non-equilibrium steady state (NESS).\\
In the current work, we extend this framework to capture the full temporal evolution of chiral active Brownian particles (ABPs) in three dimensions, encompassing the transient dynamics leading up to NESS. This approach is motivated by the universality of underlying mechanisms—steric interactions, persistent motion, and rotational diffusion—that govern density fluctuations across time scales in active matter ~\cite{Cates2015}. These interactions lead to non-equilibrium states: MIPS of ABPs in 2-dimensions and clustering in 3-dimensions, both characterized by finite memory and exponential autocorrelation decay, features well-described by the OU process. While geometric considerations (e.g., volumetric definitions of local packing) may rescale characteristic times such as the correlation time $\tau$, the underlying OU structure remains valid due to its minimal reliance on dimensionality-specific features.\\
Notably, experimental studies on microswimmer suspensions in confined three-dimensional environments suggest that active Brownian particles (ABPs) can also undergo motility-induced phase separation (MIPS) in 3D systems\cite{omar2021phase,turci2021phase}. This supports the use of our phenomenological ansatz to model the time-dependent local density 
$\rho (t)$ as an OU process, providing a minimal yet physically consistent framework for capturing its evolution across both transient and steady-state regimes. We now characterize the mean and autocorrelation properties of $\rho(t)$ under this stochastic description:
\begin{equation}
    \langle\rho(t) \rangle = (\rho_0-\rho_\infty)e^{-t/\tau} + \rho_\infty 
    \label{rho_mean}
\end{equation}
wherein $\rho_0$ is density around a local cluster which is equivalent to density in liquid phase and $\rho_\infty$ is the density in the dilute phase.
and the correlations in the density fluctuations vary as:
\begin{equation}
    \mathcal{C}_{\rho}(t_1, t_2) = \langle \delta\rho(t_1) \delta\rho(t_2) \rangle =  \sigma^2_{\rho} e^{\frac{-|t_1 - t_2|}{\tau}} 
    \label{del_rho_corr}
\end{equation}
where averaging $\langle \rangle$ is over a large number of realizations and fluctuations in local density are defined by $ \delta\rho(t)= \rho(t) - \langle \rho(t) \rangle$. The properties of $\rho(t)$ encode the essential dynamics of the active Brownian particle system, where steric interactions and activity gradually drive the continuous formation and dissolution of particle clusters in the steady state. This process is characterized by a finite memory timescale $\tau(\phi,\mathrm{Pe})$, where $\phi$ is the global packing density and $\mathrm{Pe}$ the Péclet number quantifying activity. $\tau$ physically signifies the average cluster lifetime. However, in a dilute system, the particles assume a homogeneous state with rapidly forming and breaking micro-clusters such that the average cluster lifetime and hence the density correlation time is negligible, $\tau <<  \frac{1}{D_r}$. It also implies that dilute systems achieve the non-equilibrium steady state in very short time. Local density fluctuations $\sigma_\rho^2 \propto \tau$ are also negligible. Therefore, mean-field can be applied in this limit, with $\langle\rho(t)\rangle \approx \rho_\infty$ and $\mathcal{C}_{\rho}(t_1, t_2) \approx 0$. We assume that the particle is effectively swimming in uniform density $\rho_\infty$, which is the steady-state value of local density and is directly related to the system's global packing density $\phi$. In dense systems, local density fluctuations are significant and need to be fully accounted for in form of equations \ref{rho_mean} and \ref{del_rho_corr} to calculate the effective dynamics of the particle.
\section{Analytical Calculation}
We start by solving the dynamics of the particle's orientation in equation \ref{eq:orientation_dynamics} and then proceeding to solve the coupled dynamics of position in equation \ref{eq:position_dynamics} by utilizing moments of both orientation and local density. The Langevin equation of motion for orientation $\bm{\hat{n}}$ can be simplified in terms of spherical coordinates $(\theta, \phi)$ using $\bm{\hat{n}} = (\sin\theta \cos\phi \bm{\hat{x}} + \sin\theta \sin\phi \bm{\hat{y}} + \cos\theta \bm{\hat{z}})$ as detailed in Appendix \ref{app:A}.
\begin{equation}
    \partial_t \theta = \frac{D_r}{\tan\theta} + \sqrt{2 D_r} \eta_\theta(t) 
    \label{eqn:theta}
\end{equation}

\begin{equation}
     \partial_t \phi = -\omega_0 + \frac{\sqrt{2 D_r}}{\sin\theta} \eta_\phi(t) 
     \label{eqn:phi}
\end{equation}
$\eta_\theta$ and $\eta_\phi$ are Gaussian rotational noises with zero mean and delta-correlated. The probability distribution function $P(\bm{\hat{n}})$ 
on the unit sphere can be expanded in spherical harmonics \cite{dhont1996introduction} and for a spherically symmetric chiral active Brownian particle with an axis of rotation, the leading anisotropic contribution corresponds to the   $l=2$, $m=0$. The time evolution of the  $P(\bm{\hat{n}}, t)$
 for the single-particle orientation governed by the equations of motion (\ref{eqn:theta}) and (\ref{eqn:phi}) satisfies the Fokker-Planck equation:
\begin{equation}
    \partial_t P = D_r \bm{\mathcal{R}}^2 P + \omega_0 \bm{\mathcal{R}}_z P
    \label{eqn:Fokker_planck}
\end{equation}
The Fokker-Planck equation involves the rotation operator $\bm{\mathcal{R}} \equiv \bm{\hat{n}} \times \grad_{\bm{\hat{n}}}$, which acts on the probability distribution $P( \bm{\hat{n}}, t)$ to describe how it changes under infinitesimal rotations of the particle orientation. Here, $\grad_{\bm{\hat{n}}}$ denotes the angular gradient on the unit sphere. This equation is analytically intractable for closed-form solutions of $P$. Following \cite{Pattanayak2024}, we employ a Laplace transform of Eq.(\ref{eqn:Fokker_planck}), which converts the time evolution of $P$ into an algebraic equation in $s$-space:
\begin{equation}
  -P(\bm{\hat{n}},0) + s\tilde{P}(\bm{\hat{n}}, s) = D_r \bm{\mathcal{R}}^2 \tilde{P} +\omega_0 \bm{\mathcal{R}}_z \tilde{P}
  \label{eqn:Fokker_planck_s}
\end{equation}
where $\tilde{P}(\bm{\hat{n}}, s) = \int_{0}^{\infty} e^{-st} P(\bm{\hat{n}},t)$ is the Laplace transform of $P(\bm{\hat{n}}, t)$ and we take the initial probability distribution function $P(\bm{\hat{n}}, 0) = \delta(\bm{\hat{n}} - \bm{\hat{n}}_0)$ assuming the particle starts with an orientation $\bm{\hat{n}}_0$. This equation can be used for exact calculation of moments for arbitrary functions $f(\bm{\hat{n}})$ of the orientation in $s-$space. Multiplying eqn \ref{eqn:Fokker_planck_s} by $f(\bm{\hat{n}})$ and integrating with respect to $\bm{\hat{n}}$ gives:
\begin{equation}
    -\langle f \rangle_0 + s\langle f \rangle_s = D_r\langle \bm{\mathcal{R}}^2f \rangle_s + \omega_0 \langle \bm{\mathcal{R}}_z f \rangle_s
    \label{eqn:moment}
\end{equation}
where $\langle f \rangle_0 = \int d\bm{\hat{n}} f(\bm{\hat{n}})\tilde{P}(\bm{\hat{n}}, 0)$ and $\langle f \rangle_s = \int d\bm{\hat{n}} f(\bm{\hat{n}})\tilde{P}(\bm{\hat{n}}, s)$. Our sequence of analysis would be first calculating the orientation correlation, then use it to obtain the velocity auto-correlation function and followed by the mean-squared displacement. To calculate the orientation correlation, we first calculate its moment by taking $f = \bm{\hat{n}}$ in eqn \ref{eqn:moment}. Using the properties of the rotation operator $\bm{\mathcal{R}}$, $\bm{\mathcal{R}}^2 f = -2f$ in three dimensions(as detailed in Appendix \ref{app:B}) and $\mathcal{R}_\alpha n_\beta = -\epsilon_{\alpha \beta \gamma} n_\gamma$, we solve for the moments of orientation vector in the Laplace transformed space:
\begin{equation*}
    \langle n_x \rangle_s = \frac{(s+2D_r)n_{0_x} + \omega_0 n_{0_y}}{\omega_0^2 + (s+2D_r)^2}
\end{equation*}
\begin{equation*}
    \langle n_y \rangle_s = \frac{-\omega_0 n_{0_x} + (s+2D_r)n_{0_y} }{\omega_0^2 + (s+2D_r)^2}
\end{equation*}
\begin{equation*}
    \langle n_z \rangle_s = \frac{n_{0_z}}{(s+2D_r)}
\end{equation*}
Taking inverse Laplace transform of the above expressions to obtain the moment of orientation vector as a function of time t:
\begin{equation*}
    \langle n_x (t) \rangle = e^{-2D_r t}\big[n_{0_x} \cos(\omega_0 t) + n_{0_y} \sin(\omega_0 t) \big]
    \end{equation*}
\begin{equation*}
    \langle n_y (t) \rangle = e^{-2D_r t}\big[-n_{0_x} \sin(\omega_0 t) + n_{0_y} \cos(\omega_0 t) \big] 
\end{equation*}
\begin{equation*}
    \langle n_z (t) \rangle = e^{-2D_r t} n_{0_z} 
\end{equation*}
which can be expressed in terms of the rotation operator to obtain the propagator equation:
\begin{equation}
    \langle \bm{\hat{n}}_t|x
    \bm{\hat{n}}_0 \rangle = e^{-2D_r t} \bm{\mathcal{R}}_z(\omega_0t)\bm{\hat{n}}_0
    \label{eqn:exp_val_n}
\end{equation}
Now we calculate the correlation of the particle orientation at time $t_1$ and $t_2$ using the expectation value calculated in \ref{eqn:exp_val_n}.
\begin{equation}
  \langle \bm{\hat{n}}_1 \cdot \bm{\hat{n}}_2 \rangle = \int \int \bm{\hat{n}}_1 \cdot \bm{\hat{n}}_2 P(\bm{\hat{n}}_1,t_1;\bm{\hat{n}}_2,t_2) d\bm{\hat{n}}_1 d\bm{\hat{n}}_2
  \label{eqn:acf}
\end{equation}
The joint probability distribution can be expressed as probability of transition from orientation $\bm{\hat{n}}_1$ to $\bm{\hat{n}}_2$, given $t_2>t_1$, as $P(\bm{\hat{n}}_1,t_1;\bm{\hat{n}}_2,t_2) = P(\bm{\hat{n}}_2,t_2 | \bm{\hat{n}}_1, t_1)P(\bm{\hat{n}}_1, t_1)$. Using this relation, we rewrite the correlation function in equation \ref{eqn:acf} as:
$ \langle \bm{\hat{n}}_1 \cdot \bm{\hat{n}}_2 \rangle = 
  \int \bm{\hat{n}}_1 P(\bm{\hat{n}}_1,t_1) \cdot \Big(\int \bm{\hat{n}}_2 P(\bm{\hat{n}}_2,t_2|\bm{\hat{n}}_1,t_1) d\bm{\hat{n}}_2 \Big) d\bm{\hat{n}}_1$. Using equation \ref{eqn:exp_val_n} and definition of the propagator $P(\bm{\hat{n}}_t|\bm{\hat{n}}_0, t)$ we further simplify the expression. 
\begin{equation*}
    \langle \bm{\hat{n}}_1 \cdot \bm{\hat{n}}_2 \rangle_{t_2>t_1} =
  e^{-2D_R (t_2 - t_1)} \langle \bm{\hat{n}}_1 \cdot \big [\bm{\mathcal{R}}_z(\omega_0(t_2 - t_1)\bm{\hat{n}}_1 \big] \rangle
\end{equation*}
\begin{multline}
    \langle \bm{\hat{n}}_1 \cdot \bm{\hat{n}}_2 \rangle_{t_2 > t_1} =
  e^{-2 D_R (t_2 - t_1)} \Big [\cos \omega_0(t_2 - t_1) \\ + \langle {n_1}_z^2 \rangle \big\{1 - \cos \omega_0 (t_2 - t_1)\big \} \Big]
  \label{eqn:acf_exp_1_simple}
\end{multline}
The two-time orientation auto-correlation is non-stationary as it is dependent on the orientation at time $t_1$. The second term reflects the anisotropy of the initial orientation distribution in the body frame relative to the rotation axis $\bm{\hat{z}}$. To evaluate $\langle {n_1}_z^2 \rangle$, we use the fact that the orientation distribution $P(\bm{\hat{n}}, t)$ evolves under rotational diffusion with chirality. Its second moment along $\hat{z}$ relaxes exponentially from the initial condition $n_{0z}^2$ to the isotropic value $1/3$, with a characteristic decay rate set by the $l=2$ rotational mode. This yields:
\begin{widetext}
\begin{equation}
    \langle \bm{\hat{n}}_1 \cdot \bm{\hat{n}}_2 \rangle_{t_2 > t_1} =
    e^{-2 D_R (t_2 - t_1)} \Bigg[\cos \omega_0(t_2 - t_1) + \left(\frac{1}{3} + e^{-6 D_R t_1} \left(n_{0z}^2 - \frac{1}{3}\right)\right) \big\{1 - \cos \omega_0(t_2 - t_1)\big\} \Bigg],
    \label{eqn:acf_exp_1}
\end{equation}
\end{widetext}
This decomposition follows from expanding the orientation probability density in spherical harmonics and isolating the contribution of the $l=2$, $m=0$ mode (see Ref.~\cite{Hagen2011} for similar treatment in the context of active Brownian rotation). The relaxation time of the second moment is $1/(6D_R)$, consistent with the known eigenvalue spectrum of the rotational diffusion operator in three dimensions.
\begin{figure*}[t]
    \centering
    \includegraphics[width=0.80\textwidth]{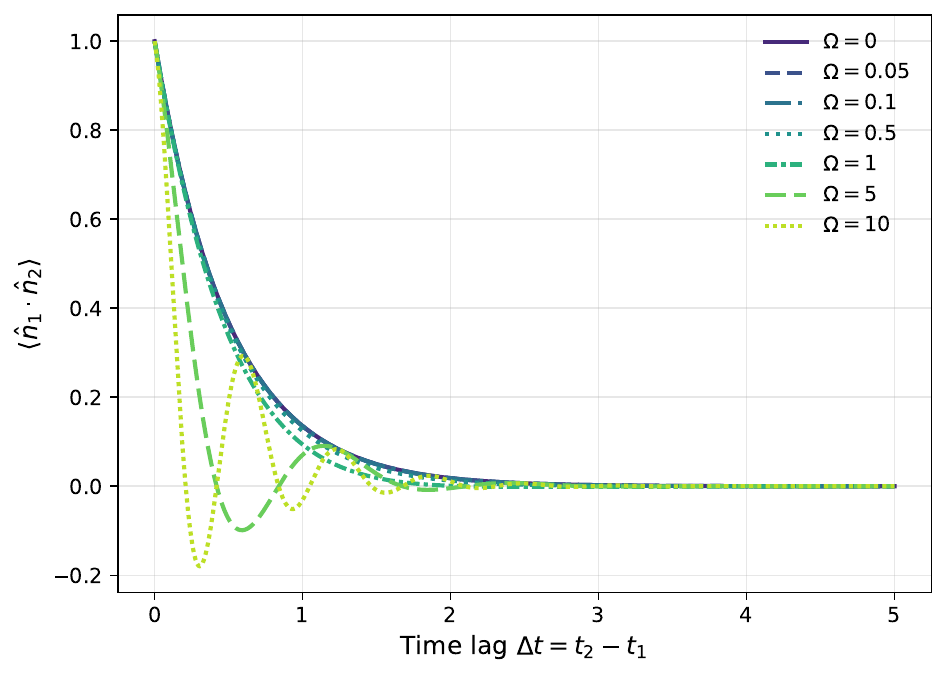}
    \caption{Temporal orientation correlations $\langle \hat{\mathbf{n}}_1 \cdot \hat{\mathbf{n}}_2 \rangle$ for a 3D chiral active Brownian particle, showing: 
(a)~Dependence on dimensionless chirality $\omega_0/D_r$ at fixed $t_1=0.5$, $n_{0z}=0.85$; 
The analytical solution (solid curves) from Eq.~\eqref{eqn:acf_exp_1} exhibits characteristic features: 
(i)~Exponential decay $e^{-2D_r\Delta t}$ (dashed gray), 
(ii)~Chiral oscillations with frequency $\omega_0$, and 
(iii)~Non-Markovian dependence on both $\Delta t \equiv t_2-t_1$ and $t_1$. 
The $n_{0z}$-dependent prefactor $[\frac{1}{3} + e^{-6D_r t_1}(n_{0z}^2 - \frac{1}{3})]$ modulates oscillation amplitudes, showing memory of initial conditions. 
All plots use $D_r=1$ as the fundamental timescale.}
    \label{fig:corr_orient}
\end{figure*}
The orientation correlation function is, in general, sensitive to the initial orientation component $n_{0z}$; however, upon averaging over uniformly distributed initial orientations, for which $\langle {n_0}_z^2 \rangle = 1/3$, the non-stationary term depending on absolute time $t_1$ in \ref{eqn:acf_exp_1} vanishes. We thus employ the stationary form of the orientation correlation for calculating the MSD averaged over all possible initial orientations of the particle.
\begin{equation}
     \langle \bm{\hat{n}}_1 \cdot \bm{\hat{n}}_2 \rangle = \mathcal{C}_{\bm{\hat{n}}}(\Delta t) = \frac{1}{3} e^{-2D_r |t_2 - t_1|} \big(1+ 2 \cos \omega_0 |t_2 - t_1| \big) 
     \label{eqn: acf_orientation_avgd}
\end{equation}
The behavior of $\langle \bm{\hat{n}}_1 \cdot \bm{\hat{n}}_2 \rangle$ for different cases in shown in Figure \ref{fig:corr_orient}. 
The mean-squared displacement (MSD) of the chiral active Brownian particle (cABP) follows from the Langevin equation (Eq.~(1)) via the Green-Kubo relation:
\begin{equation}
\Delta (t) = \langle |\bm{r}(t) - \bm{r}(0)|^2 \rangle =  \int_0^t \int_0^{t} \langle \bm{v}(t_1) \cdot \bm{v}(t_2) \rangle \, dt_1 \, dt_2,
\label{eq:MSD}
\end{equation}
where the velocity autocorrelation function (VACF) has contributions from local density fluctuations, rotational and translational noise as in Langevin equation \ref{eqn:pos}. The VACF can be further expressed as the product of speed and orientation correlations, since both are independent stochastic variables:
\begin{equation}
\langle \bm{v}(t_1) \cdot \bm{v}(t_2) \rangle = \langle v_\rho(t_1) v_\rho(t_2) \rangle \mathcal{C}_{\bm{\hat{n}}}(\Delta t) + 6D \delta (t_1 - t_2)
\label{eq:VACF}
\end{equation}
Now we calculate the VACF in the dilute and dense limits:
\subsection{Dilute Regime ($\phi < \phi_c$)}
In the dilute limit, micro-clusters form and break rapidly ($\tau \to 0$) to achieve a homogeneous state, and density fluctuations are negligible at all times. As discussed previously, we take the approximations $ \langle\rho(t)\rangle  \approx \rho_\infty$ and $\langle \delta\rho(t_1)\delta\rho(t_2) \rangle \approx 0$ to obtain the speed auto-correlation in the dilute limit :
\begin{equation*}
\langle v_\rho(t_1) v_\rho(t_2) \rangle \approx v_0^2 \left(1 - 2\lambda \rho_\infty + \lambda^2 \rho_\infty^2\right) = v_0^2 (1 - \lambda \rho_\infty)^2,
\label{eq:speed_dilute_approx}
\end{equation*}
Substituting the above value in Eq.\ref{eq:VACF} we get:
\begin{equation}
\langle \bm{v}(t_1) \cdot \bm{v}(t_2) \rangle = v_0^2 (1 - \lambda \rho_\infty)^2  \mathcal{C}_{\bm{\hat{n}}}(t_2 - t_1) + 6D \delta (t_2 - t_1) 
\label{eq:VACF_dilute}
\end{equation}
Using relation \ref{eq:MSD}, \ref{eq:VACF} and \ref{eq:VACF_dilute} to calculate mean-squared displacement (MSD) from velocity auto-correlation function

\begin{widetext}

    \begin{align*}
    \langle \bm{r}^2 \rangle &= 6Dt +  v_0^2 (1 - \lambda \rho_\infty)^2 \int_0^t \int_0^{t} \mathcal{C}_{\bm{\hat{n}}}(\Delta t) dt_1 \, dt_2,\\
        \langle \bm{r}^2 \rangle &= 6Dt +  v_0^2 (1 - \lambda \rho_\infty)^2 \int_0^t \int_0^{t}\frac{1}{3} e^{-2D_r |t_2 - t_1|} \big(1+ 2 \cos \omega_0 |t_2 - t_1| \big) dt_1 \, dt_2,\\ 
        \end{align*}
        \begin{align*}
     \langle \bm{r}^2 \rangle = 6Dt + \frac{v_0^2}{6D_r^2} (1 - \lambda \rho_\infty)^2 \Big\{2D_r t + e^{-2D_r t} - 1\Big\} &+ \frac{4v_0^2}{3(4D_r^2 + \omega_0^2)^2} (1 - \lambda \rho_\infty)^2 \Big\{\omega_0^2 - 4D_r^2 \\ 
     &+ 2D_r(4D_r^2 + \omega_0^2)t + e^{-2D_r t}\big( (4D_r^2 - \omega_0^2)\cos \omega_0 t  - 4\omega_0 D_r \sin \omega_0 t\big) \}
        \end{align*}
    Introducing measures of activity and chirality as dimensionless constants $Pe = \frac{v_0}{\sqrt{D D_r}}$ and $\Omega = \frac{\omega_0}{D_r}$. Expressing the MSD in dimensionless form using the dimensionless units of time $\tilde{t} = D_r t$ and position $\tilde{r} = r/\sqrt{\frac{D}{D_r}}$
    \begin{align*}
    \langle \bm{\tilde{r}}^2 \rangle &= 6\tilde{t} + \frac{Pe^2}{6} (1 - \lambda \rho_\infty)^2 \Big\{2\tilde{t} + e^{-2\tilde{t}} - 1\Big\} \\ &+ \frac{64}{3}\frac{Pe^2}{(\Omega^2 + 4)^2} (1 - \lambda \rho_\infty)^2 \Big\{\Omega^2 - 4 + 2(\Omega^2 + 4)\tilde{t} + e^{-2 \tilde{t}}\big( (4 - \Omega^2)\cos \Omega t  - 4\Omega \sin \Omega t\big) \Big\} \numberthis 
    \label{MSD:dilute}
    \end{align*}
    The above eqn for $\tilde{t}\to\infty$ yields
    \begin{align*}
    \langle \bm{\tilde{r}}^2 \rangle &=
    6\tilde{t}\Big\{1+\frac{Pe^2}{18}(1-\lambda\rho_\infty)^2+\frac{64}{9}\frac{Pe^2}{(\Omega^2+4)}(1-\lambda\rho_\infty)^2\Big\}
    \end{align*}
    \qquad where, $D_{\mathrm{eff}}=1+\frac{Pe^2}{18}(1-\lambda\rho_\infty)^2+\frac{64}{9}\frac{Pe^2}{(\Omega^2+4)}(1-\lambda\rho_\infty)^2$. 
    In the long-time limit ($\tilde{t} \to \infty$), the MSD reduces to a linear growth with an effective diffusivity
\[
D_{\mathrm{eff}} = 1 + \frac{Pe^2}{18}(1-\lambda \rho_\infty)^2 + \frac{64}{9}\frac{Pe^2}{\Omega^2+4}(1-\lambda \rho_\infty)^2.
\]
This expression exhibits a quadratic dependence on activity ($Pe^2$) and a suppression of diffusivity with increasing chirality ($\Omega$), consistent with the functional form obtained for a single chiral Brownian swimmer in \cite{vanTeeffelen2008}, where the effective diffusivity scales as $v_0^2/(D_r^2 + \Omega^2)$.
    In the achiral limit ,$\Omega=0$ we see that \ref{MSD:dilute} reduces to
    \begin{align*}
     \langle \bm{\tilde{r}}^2 \rangle &=
     6\tilde{t} +\frac{11 Pe^2}{2}(1-\lambda\rho_\infty)^2\Big\{2\tilde{t}+e^{-2\tilde{t}}-1\Big\}  \tag{20A}
     \label{eq:msd_achiral}
    \end{align*}
    (a)For $\tilde{t}\gg1$:\\
    \begin{align*}
\langle \bm{\tilde{r}}^2 \rangle
&=
6\tilde{t}\Big\{1+\frac{11}{6}Pe^2(1-\lambda\rho_\infty)^2\Big\},\tag{20B}\
\text{where }
D_{\mathrm{eff}}
=
1+\frac{11}{6}Pe^2(1-\lambda\rho_\infty)^2 .
\end{align*}
(b)For $\tilde{t}\ll 1$
\begin{align*}
\langle \bm{\tilde{r}}^2 \rangle =
6\tilde{t}+(1-\lambda\rho_\infty)^211Pe^2\tilde{t}^2 \tag{20C}
\end{align*}

The expression of MSD for low density  and zero chirality limit as shown in Eq.\ref{eq:msd_achiral}, is consistent with previously reported calculation of MSD of ABP in three dimensions in \cite{kurzthaler2016intermediate}. In particular, it shows the characteristic crossover from short-time ballistic motion to long-time diffusive behavior, as well as the activity-induced enhancement of the effective diffusivity.

\end{widetext} 
\subsection{Dense Regime ($\phi > \phi_c$)}
In dense systems, local density fluctuation are significant and mean-field approximation does not work. Thus, we retain the full expressions for $\langle \rho(t) \rangle$ as in equation \ref{rho_mean} and correlations in equation \ref{del_rho_corr} while calculating the speed auto-correlation. For analytical tractability, we perform a second-order truncation in the local density fluctuations, retaining contributions up to the level of the mean density and the two-point correlation function $C_\rho(t_1,t_2)$. This corresponds to a lowest-order closure in which higher-order cumulants (such as three- and four-point correlations) are neglected. We emphasize that this approximation should be viewed as a minimal coarse-grained description that captures the leading order contributions in local density fluctuations in the system.
. Since $\rho(t)$ follows an Ornstein–Uhlenbeck process with non-zero mean and finite correlation time $\tau$, we expand the effective speed function $v_\rho(t)$ to second order in $\rho$, and keep terms up to first order in $\langle \rho(t) \rangle$ and $\mathcal{C}_\rho(t_1, t_2) = \langle \delta \rho(t_1) \delta \rho(t_2) \rangle$. This approximation captures the dominant corrections arising from density-induced motility suppression, while neglecting subleading higher-order (e.g., three- and four-point) correlations that contribute weakly to the MSD. Physically, the linear correction terms reflect the coupling of transient density buildup to the slowing of active motion, whereas the two-point correlation encodes finite memory effects in speed fluctuations due to spatial clustering. Thus, we get the speed auto-correlation function with two correction terms:

\begin{widetext}
\begin{equation}
\begin{aligned}
\langle v_\rho(t_1) v_\rho(t_2) \rangle 
&\approx v_0^2 \Bigl[1
- \underbrace{\lambda_1 \bigl\{1 + 2 \lambda_2 \,\mathcal{C}_\rho(t_1, t_2)\bigr\}
      \bigl(\langle \rho_1\rangle + \langle \rho_2 \rangle\bigr)
}_{\text{T: transient local density}}
+ \underbrace{\lambda_1^2 \,\mathcal{C}_\rho(t_1, t_2)}_{\text{S: steady-state fluctuations}}
\Bigr]
\end{aligned}
\label{eq:speed_dense_full}
\end{equation}
Using relations \ref{eq:MSD}\, \ref{eq:VACF} and \ref{eq:speed_dense_full}, we obtain the velocity auto-correlation function of the particle and then calculate the MSD of the particle in dense regime.
 All fourth order correlations are ignored here. This corresponds to a second order closure in density fluctuations, which is expected to be valid when the variance $\sigma_\rho^2$ is moderate and higher-order cumulants provide subleading corrections.

\begin{equation}
\begin{aligned}
\langle \bm{r}^2 \rangle 
&= 6Dt + v_0^2 \int_0^t\int_0^t \Bigl[1
- \lambda_1 \bigl\{1 + 2 \lambda_2 \,\mathcal{C}_\rho(\Delta t)\bigr\}
      \bigl(\langle \rho_1\rangle + \langle \rho_2 \rangle\bigr)
+ \lambda_1^2 \,\mathcal{C}_\rho(\Delta t)
\Bigr]\mathcal{C}_{\bm{\hat{n}}}(\Delta t)dt_1 \, dt_2\\ 
\end{aligned}
\end{equation}
Where $\Delta t = t_1 - t_2 $ and each term is an integral transform of the functions or a combination of the product of the the following functions: orientation auto-correlation function $\langle \bm{\hat{n}}_1 \cdot\bm{\hat{n}}_2  \rangle = \mathcal{C}_{\bm{\hat{n}}}(\Delta t)$, density correlation function $\mathcal{C}_\rho(\Delta t)$ and mean density $\langle \rho_t \rangle$, with values given in equation \ref{eqn: acf_orientation_avgd}, \ref{del_rho_corr} and \ref{rho_mean} . We define the following operators to express these integrals compactly:

\begin{align*}
\mathcal{I}[\mathcal{C}_{\bm{\hat{n}}}](t) &\equiv \int_0^t \int_0^t \mathcal{C}_{\bm{\hat{n}}}(\Delta t) dt_1 dt_2, \\
\mathcal{I}[\mathcal{C}_{\bm{\hat{n}}, \beta}](t) &\equiv \int_0^t \int_0^t \mathcal{C}_\rho(\Delta t) \mathcal{C}_{\bm{\hat{n}}}(\Delta t) dt_1 dt_2,  \\
\mathcal{F}_\tau[\mathcal{C}_{\bm{\hat{n}}}](t) &\equiv \int_0^t \int_0^t e^{-t_1/\tau}\mathcal{C}_{\bm{\hat{n}}}(\Delta t) dt_1 dt_2, \\
\mathcal{F}_\tau[\mathcal{C}_{\bm{\hat{n}}, \beta}](t) &\equiv \int_0^t \int_0^t e^{-t_1/\tau}\mathcal{C}_\rho(\Delta t) \mathcal{C}_{\bm{\hat{n}}}(\Delta t) dt_1 dt_2
\end{align*}
$\mathcal{I}[\mathcal{C}_{\bm{\hat{n}}}]$ represents integral over (square time-domain) of the symmetric orientation autocorrelation $\mathcal{C}_{\bm{\hat{n}}}(\Delta t) = \frac{1}{3} e^{-2D_r |\Delta t|}(1 + 2\cos \omega_0 |\Delta t|)$, $\mathcal{I}[\mathcal{C}_{\bm{\hat{n}}, \beta}]$ represents integral of the product of orientation and local density correlations $\mathcal{C}_{\bm{\hat{n}},\beta}(\Delta t) = \mathcal{C}_{\bm{\hat{n}}}(\Delta t) \mathcal{C}_{\rho}(\Delta t)  = \frac{1}{3} e^{-(2D_r + \frac{1}{\tau}) |\Delta t|}(1 + 2\cos \omega_0 |\Delta t|) = \frac{1}{3} e^{-\beta |\Delta t|}(1 + 2\cos \omega_0 |\Delta t|)$, resulting in effective relaxation timescale  $\beta = 2D_r + \frac{1}{\tau}$ which couples local density fluctuations in spatial clusters in the non-equilibrium steady state to the effective spatial dynamics of the particle.  
$\mathcal{F}_\tau[\mathcal{C}_{\bm{\hat{n}}}]$ is the integral transform of the orientation correlation weighted by an exponential memory kernel, 
$\mathcal{K}(t_1) = e^{-t_1/\tau}$ (or equivalently $e^{-t_2/\tau}$), 
describing the spatial dynamics of the particle during the transient regime in which local density around is rapidly re-organizing. 
$\mathcal{F}_\tau[\mathcal{C}_{\bm{\hat{n}}, \beta}]$ similarly captures spatial dynamics of the particle in transient local density with correlated density fluctuations, which drives the kinetic growth of micro-clusters. 
Although these features decay over time, they leave a finite imprint on the dynamics due to the finite memory encoded in $\tau$. These operators cleanly isolate the physically distinct contributions to the mean-squared displacement: 
$\mathcal{I}[\mathcal{C}_{\bm{\hat{n}}}]$ governs the baseline diffusive enhancement due to persistent self-propulsion, 
$\mathcal{F}_\tau[\cdot]$ encodes the influence of transient local density buildup, 
and $\mathcal{I}[\mathcal{C}_{\bm{\hat{n}},\beta}]$ captures the persistent fluctuations arising from spatially structured clusters.

\begin{align}
\langle \bm{r}^2(t) \rangle= 6Dt + v_0^2 \bigg[(1 - 2\lambda_1 \rho_\infty)\, \mathcal{I}[\mathcal{C}_{\bm{\hat{n}}}](t) 
+ 2\lambda_1 (\rho_\infty - \rho_0)\big\{\mathcal{F}_\tau[\mathcal{C}_{\bm{\hat{n}}}](t) + 2\lambda_2 \sigma_\rho^2 \mathcal{F}_\tau[\mathcal{C}_{\bm{\hat{n}}, \beta}](t)\big\}
+  \lambda_1(\lambda_1 - 4\lambda_2 \rho_\infty) \sigma_\rho^2\, \mathcal{I}[\mathcal{C}_{\bm{\hat{n}}, \beta}](t) \bigg]
\label{msd_opr_form}
\end{align}
The base integral $\mathcal{I}[\mathcal{C}_{\bm{\hat{n}}}](t)$ has already been evaluated in Eq.~(20) for the dilute case. The integral and $\mathcal{I}[\mathcal{C}_{{\bm{\hat{n}}},\beta}](t)$ is computed analogously replacing $D_r \xrightarrow{} \beta = D_r + 1/\tau$, and yield closed-form expressions involving damped oscillations. We evaluated both $\mathcal{F}_\tau[\mathcal{C}_{\bm{\hat{n}},\beta}]$ and $\mathcal{F}_\tau[\mathcal{C}_{\bm{\hat{n}}, \beta}]$ using the convolution identity:
\begin{equation}
\mathcal{F}_\tau[\mathcal{C}_{\bm{\hat{n}}}](t)
=
\int_0^t\int_0^t e^{-t_1/\tau}\,\mathcal{C}_{\bm{\hat{n}}}(t_2-t_1)\,dt_1dt_2
\;=\;
\mathcal{I}[\mathcal{C}_{\bm{\hat{n}}}](t)
\;-\;
\frac{1}{\tau}\int_0^t e^{-(t-s)/\tau}\,\mathcal{I}[\mathcal{C}_{\bm{\hat{n}}}](s)\,ds
\label{memory_operator}
\end{equation}
and introduce the shorthand memory‐operator
$
\mathcal{T}_\tau[f](t)\;\equiv\;\frac{1}{\tau}\int_0^t e^{-(\,t-s)/\tau}\,f(s)\,ds.
$
Hence 
\(\mathcal{F}_\tau=\mathcal{I}-\mathcal{T}_\tau\circ\mathcal{I}\), and 
\begin{align}
  \mathcal{T}_\tau[\mathcal{C}_{\bm{\hat{n}}}](t)\; =  \frac{v_0^2}{6D_r^2}   \Big\{2D_r \tau (1 - e^{-\frac{t}{\tau}}) - \frac{e^{-2D_r t} - e^{-\frac{t}{\tau}}}{2D_r-\frac{1}{\tau}}\Big\} 
     + \frac{4v_0^2}{3(4D_r^2 + \omega_0^2)^2}  \Big\{A_0 (1 - e^{-\frac{t}{\tau}}) + B_0 \frac{e^{-2 D_r t} - e^{-\frac{t}{\tau}}}{\frac{1}{\tau^2} + \omega_0^2} \Big \}
\end{align}
\begin{align}
    \mathcal{T}_\tau[\mathcal{C}_{\bm{\hat{n}}, \beta}](t)\; = \frac{2v_0^2}{3\beta^2}\Big\{\beta \tau (1 - e^{-\frac{t}{\tau}}) - \frac{e^{-\beta t} - e^{-\frac{t}{\tau}}}{\beta-\frac{1}{\tau}} \Big\} 
     + \frac{4v_0^2}{3(\beta^2 + \omega_0^2)^2} \Big\{A_\beta (1 - e^{-\frac{t}{\tau}}) + B_\beta \frac{e^{-\beta t} - e^{-\frac{t}{\tau}}}{\frac{1}{\tau^2} + \omega_0^2} \Big\} 
\end{align}
where \( A_0 = \dfrac{\omega_0^2 - 4D_r^2}{1/\tau} + \dfrac{2D_r(4D_r^2 + \omega_0^2)}{(1/\tau)^2 + \omega_0^2} \), \quad
\( B_0 = (4D_r^2 - \omega_0^2)\, \dfrac{1}{\tau} - 4D_r \omega_0 \) and \( A_\beta = \dfrac{\omega_0^2 - \beta^2}{1/\tau} + \dfrac{\beta(2\beta^2 + \omega_0^2)}{(1/\tau)^2 + \omega_0^2} \), \\ \quad
\( B_\beta = (\beta^2 - \omega_0^2)\, \dfrac{1}{\tau} - 2\beta \omega_0 \)\\
Using relations \ref{memory_operator} in \ref{msd_opr_form} simplifies MSD expression to a form that demonstrating that the particle does not fully lose memory of its initial local environment: the long-time diffusivity carries an imprint of the starting density $\rho_0$, due to presence if nonequlibrium phase separation in the system and  the non-ergodic exploration of phase space across coexisting dense and dilute basins.
\begin{align}
\langle \bm{r}^2(t) \rangle= 6Dt + v_0^2 \bigg[(1 - 2\lambda_1 \rho_0)\, \mathcal{I}[\mathcal{C}_{\bm{\hat{n}}}](t) 
- 2\lambda_1 (\rho_\infty - \rho_0)\big\{\mathcal{T}_\tau[\mathcal{C}_{\bm{\hat{n}}}](t) + 2\lambda_2 \sigma_\rho^2 \mathcal{T}_\tau[\mathcal{C}_{\bm{\hat{n}}, \beta}](t)\big\}
+  \lambda_1(\lambda_1 - 4\lambda_2 \rho_0) \sigma_\rho^2\, \mathcal{I}[\mathcal{C}_{\bm{\hat{n}}, \beta}](t) \bigg]
\label{msd_opr_form_simplfied}
\end{align}
Substituting all contributions, we obtain the full expression for the MSD of a chiral ABP in the dense regime, as a function of initial and steady-state density parameters.The details of the calculation are as provided in Appendix \ref{app:C}.

 \begin{align}
     \langle \bm{r}^2 \rangle &= 6Dt + \frac{v_0^2}{6D_r^2} (1 - 2\lambda_1 \rho_0) \Big\{2D_r t + e^{-2D_r t} - 1\Big\} \nonumber\\
     &+ \frac{4v_0^2}{3(4D_r^2 + \omega_0^2)^2} (1 - 2\lambda_1 \rho_0)\Big\{\omega_0^2 - 4D_r^2 + 2D_r(4D_r^2 + \omega_0^2)t + e^{-2D_r t}\big( (4D_r^2 - \omega_0^2)\cos \omega_0 t  - 4\omega_0 D_r \sin \omega_0 t\big) \Big\} \nonumber\\
    &+ \frac{2v_0^2}{3\beta^2}\sigma_\rho^2 \lambda_1(\lambda_1 - 4\lambda_2 \rho_0) \Big\{\beta t + e^{-\beta t} - 1\Big\} \nonumber\\
     &+ \frac{4v_0^2}{3(\beta^2 + \omega_0^2)^2}\sigma_\rho^2  \lambda_1(\lambda_1 - 4\lambda_2 \rho_0)\Big\{\omega_0^2 - \beta^2 + \beta(\beta^2 + \omega_0^2)t + e^{-\beta t}\big( (\beta^2 - \omega_0^2)\cos \omega_0 t  - 2\omega_0 \beta \sin \omega_0 t\big) \Big\} \nonumber \\
     &- \frac{v_0^2}{6D_r^2} \cdot 2\lambda_1 (\rho_\infty - \rho_0) \Big\{2D_r \tau (1 - e^{-\frac{t}{\tau}}) - \frac{e^{-2D_r t} - e^{-\frac{t}{\tau}}}{2D_r-\frac{1}{\tau}}\Big\} \nonumber\\
     &- \frac{4v_0^2}{3(4D_r^2 + \omega_0^2)^2} \cdot 2\lambda_1 (\rho_\infty - \rho_0)\Big\{A_0 (1 - e^{-\frac{t}{\tau}}) + B_0 \frac{e^{-2D_r t} - e^{-\frac{t}{\tau}}}{\frac{1}{\tau^2} + \omega_0^2} \Big \} \nonumber\\
    &- \frac{2v_0^2}{3\beta^2}\sigma_\rho^2\cdot 4\lambda_1 \lambda_2(\rho_\infty - \rho_0) \Big\{\beta \tau (1 - e^{-\frac{t}{\tau}}) - \frac{e^{-\beta t} - e^{-\frac{t}{\tau}}}{\beta-\frac{1}{\tau}} \Big\} \nonumber\\
     &- \frac{4v_0^2}{3(\beta^2 + \omega_0^2)^2}\sigma_\rho^2 \cdot 4\lambda_1 \lambda_2(\rho_\infty - \rho_0)\Big\{A_\beta (1 - e^{-\frac{t}{\tau}}) + B_\beta \frac{e^{-\beta t} - e^{-\frac{t}{\tau}}}{\frac{1}{\tau^2} + \omega_0^2} \Big\} 
     \label{msd_dense}
        \end{align}
To interpret the full MSD expression in terms of dimensionless activity and chirality, we define nondimensional variables: $\tilde{t} = D_r t$, $\Omega = \omega_0 / D_r$, $\text{Pe} = v_0 / \sqrt{D D_r}$, and $\tilde{\tau} = D_r \tau$. The MSD is scaled by the passive Brownian unit $\langle \tilde{\bm{r}}^2 \rangle = \langle \bm{r}^2 \rangle / (D / D_r)$, yielding:

\begin{align*}
\langle \tilde{\bm{r}}^2(\tilde{t}) \rangle &= 6\tilde{t} + \frac{\text{Pe}^2}{6}(1 - 2\lambda_1 \rho_0)\left\{2\tilde{t} + e^{-2\tilde{t}} - 1 \right\} \\
&+ \frac{4\,\text{Pe}^2}{3(4 + \Omega^2)^2}(1 - 2\lambda_1 \rho_0)\left\{ \Omega^2 - 4 + 2(4 + \Omega^2)\tilde{t} + e^{-2\tilde{t}} \left[ (4 - \Omega^2)\cos(\Omega \tilde{t}) - 4\Omega \sin(\Omega \tilde{t}) \right] \right\} \\
&+ \frac{2\,\text{Pe}^2}{3\tilde{\beta}^2} \sigma_\rho^2 \lambda_1(\lambda_1 - 4\lambda_2 \rho_0)\left\{ \tilde{\beta}\tilde{t} + e^{-\tilde{\beta} \tilde{t}} - 1 \right\} \\
&+ \frac{4\,\text{Pe}^2}{3(\tilde{\beta}^2 + \Omega^2)^2} \sigma_\rho^2 \lambda_1(\lambda_1 - 4\lambda_2 \rho_0)\left\{ \Omega^2 - \tilde{\beta}^2 + \tilde{\beta}(\tilde{\beta}^2 + \Omega^2)\tilde{t} + e^{-\tilde{\beta} \tilde{t}} \left[ (\tilde{\beta}^2 - \Omega^2)\cos(\Omega \tilde{t}) - 2\tilde{\beta} \Omega \sin(\Omega \tilde{t}) \right] \right\} \\
&- \frac{\text{Pe}^2}{6} \cdot 2\lambda_1(\rho_\infty - \rho_0) \left\{2\tilde{\tau}(1 - e^{-\tilde{t}/\tilde{\tau}}) - \frac{e^{-2\tilde{t}} - e^{-\tilde{t}/\tilde{\tau}}}{2 - \frac{1}{\tilde{\tau}}} \right\} \\
&- \frac{4\,\text{Pe}^2}{3(4 + \Omega^2)^2} \cdot 2\lambda_1(\rho_\infty - \rho_0)\left\{ \tilde{A}_0(1 - e^{-\tilde{t}/\tilde{\tau}}) + \tilde{B}_0 \frac{e^{-2\tilde{t}} - e^{-\tilde{t}/\tilde{\tau}}}{1/\tilde{\tau}^2 + \Omega^2} \right\} \\
&- \frac{2\,\text{Pe}^2}{3\tilde{\beta}^2} \sigma_\rho^2 \cdot 4\lambda_1 \lambda_2 (\rho_\infty - \rho_0) \left\{ \tilde{\beta} \tilde{\tau} (1 - e^{-\tilde{t}/\tilde{\tau}}) - \frac{e^{-\tilde{\beta} \tilde{t}} - e^{-\tilde{t}/\tilde{\tau}}}{\tilde{\beta} - \frac{1}{\tilde{\tau}}} \right\} \\
&- \frac{4\,\text{Pe}^2}{3(\tilde{\beta}^2 + \Omega^2)^2} \sigma_\rho^2 \cdot 4\lambda_1 \lambda_2 (\rho_\infty - \rho_0)\left\{ \tilde{A}_\beta(1 - e^{-\tilde{t}/\tilde{\tau}}) + \tilde{B}_\beta \frac{e^{-\tilde{\beta} \tilde{t}} - e^{-\tilde{t}/\tilde{\tau}}}{1/\tilde{\tau}^2 + \Omega^2} \right\}
\tag{28A}
\label{msd_dense_reduced}
\end{align*}

where the dimensionless chirality-coupled prefactors are:
\begin{align*}
\tilde{A}_0 &= \frac{\Omega^2 - 4}{1/\tilde{\tau}} + \frac{2(4 + \Omega^2)}{1/\tilde{\tau}^2 + \Omega^2}, \quad
\tilde{B}_0 = (4 - \Omega^2)\cdot \frac{1}{\tilde{\tau}} - 4\Omega, \\
\tilde{A}_\beta &= \frac{\Omega^2 - \tilde{\beta}^2}{1/\tilde{\tau}} + \frac{\tilde{\beta}(2\tilde{\beta}^2 + \Omega^2)}{1/\tilde{\tau}^2 + \Omega^2}, \quad
\tilde{B}_\beta = (\tilde{\beta}^2 - \Omega^2)\cdot \frac{1}{\tilde{\tau}} - 2\tilde{\beta} \Omega.
\end{align*}
Eq.\ref{msd_dense_reduced} gives the mean-squared displacement conditioned on a specified initial local density $\rho_0$. If instead the initial condition is drawn from the stationary density ensemble, so that $\rho_0=\rho_\infty$, all transient contributions proportional to $(\rho_\infty-\rho_0)$ vanish identically. In this case, the MSD reduces to a purely steady-state form, with long-time transport characterized by an effective diffusivity $D_\infty$ independent of the initial density basin. For completeness, we provide below the dense-regime MSD evaluated for $\rho_0=\rho_\infty$.
\begin{align*}
\langle \tilde{\bm{r}}^2(\tilde{t}) \rangle &= 6\tilde{t} + \frac{\text{Pe}^2}{6}(1 - 2\lambda_1 \rho_0)\left\{2\tilde{t} + e^{-2\tilde{t}} - 1 \right\} \\
&+ \frac{4\,\text{Pe}^2}{3(4 + \Omega^2)^2}(1 - 2\lambda_1 \rho_0)\left\{ \Omega^2 - 4 + 2(4 + \Omega^2)\tilde{t} + e^{-2\tilde{t}} \left[ (4 - \Omega^2)\cos(\Omega \tilde{t}) - 4\Omega \sin(\Omega \tilde{t}) \right] \right\} \\
&+ \frac{2\,\text{Pe}^2}{3\tilde{\beta}^2} \sigma_\rho^2 \lambda_1(\lambda_1 - 4\lambda_2 \rho_0)\left\{ \tilde{\beta}\tilde{t} + e^{-\tilde{\beta} \tilde{t}} - 1 \right\} \\
&+ \frac{4\,\text{Pe}^2}{3(\tilde{\beta}^2 + \Omega^2)^2} \sigma_\rho^2 \lambda_1(\lambda_1 - 4\lambda_2 \rho_0)\left\{ \Omega^2 - \tilde{\beta}^2 + \tilde{\beta}(\tilde{\beta}^2 + \Omega^2)\tilde{t} + e^{-\tilde{\beta} \tilde{t}} \left[ (\tilde{\beta}^2 - \Omega^2)\cos(\Omega \tilde{t}) - 2\tilde{\beta} \Omega \sin(\Omega \tilde{t}) \right] \right\}
\tag{28B}
\label{msd_den_avg}
\end{align*} 
This compact yet complete form allows direct comparison across parameter regimes of interest — particularly in examining the role of initial density, activity, chirality, and memory timescale — while revealing the crossover from transient enhancement to steady-state suppression in dense active systems.
\subsubsection*{Early- and late-time asymptotic behavior of MSD}

The dimensionless mean-squared displacement in the dense regime admits a well-defined asymptotic expansion in both the early- and late-time limits. In the early-time regime \( \tilde{t} \ll 1 \), all exponential and trigonometric functions in the full expression of Eq.\ref{msd_dense_reduced} can be expanded in Taylor series, yielding
\begin{equation}
\begin{aligned}
\langle \tilde{\bm{r}}^2(\tilde{t}) \rangle =\; & \big[6 -\text{Pe}^2 \lambda_1 (\rho_\infty - \rho_0)\,f(\Omega, \tilde{\tau}, \sigma_\rho) \big]\,\tilde{t} + \text{Pe}^2\, g(\Omega, \tilde{\tau}, \sigma_\rho)\, \tilde{t}^2,
\end{aligned}
\label{eq:msd_short_time}
\end{equation}
where \( f \) and \( g \) are specific functions of the system parameters $\Omega, \tilde{\tau}, \sigma_\rho^2 $ as expressed below

\begin{align*}
f(\Omega,\tilde{\tau},\sigma_\rho)
=\;& 1
+\frac{8}{3(4+\Omega^2)^2}
\left[
\frac{\tilde{A}_0}{\tilde{\tau}}
-\frac{\tilde{B}_0\left(2-\frac{1}{\tilde{\tau}}\right)}
{\frac{1}{\tilde{\tau}^2}+\Omega^2}
\right]
+\frac{8\,\sigma_\rho^2\,\lambda_2}{3\tilde{\beta}^2}\,(\tilde{\beta}+1)
\nonumber\\
&\quad
+\frac{16\,\sigma_\rho^2\,\lambda_2}{3(\tilde{\beta}^2+\Omega^2)^2}
\left[
\frac{\tilde{A}_\beta}{\tilde{\tau}}
-\frac{\tilde{B}_\beta\left(\tilde{\beta}-\frac{1}{\tilde{\tau}}\right)}
{\frac{1}{\tilde{\tau}^2}+\Omega^2}
\right].
\end{align*}
and
\begin{align*}
g(\Omega,\tilde{\tau},\sigma_\rho)
=\;& (1-2\lambda_1\rho_0)
+\sigma_\rho^2\,\lambda_1(\lambda_1-4\lambda_2\rho_0)
+\frac{\lambda_1(\rho_\infty-\rho_0)}{3}
\left[
\frac{1}{\tilde{\tau}}
+\frac{\left(2-\frac{1}{2\tilde{\tau}^2}\right)}
{2-\frac{1}{\tilde{\tau}}}
\right]
\nonumber\\
&\quad
-\frac{8\,\lambda_1(\rho_\infty-\rho_0)}{3(4+\Omega^2)^2}
\left[
-\frac{\tilde{A}_0}{2\tilde{\tau}^2}
+\frac{\tilde{B}_0\left(2-\frac{1}{2\tilde{\tau}^2}\right)}
{\frac{1}{\tilde{\tau}^2}+\Omega^2}
\right]
\nonumber\\
&\quad
-\frac{8\,\sigma_\rho^2\,\lambda_1\lambda_2(\rho_\infty-\rho_0)}{3\tilde{\beta}^2}
\left[
-\frac{\tilde{\beta}}{2\tilde{\tau}}
-\frac{\left(\frac{\tilde{\beta}^2}{2}-\frac{1}{2\tilde{\tau}^2}\right)}
{\tilde{\beta}-\frac{1}{\tilde{\tau}}}
\right]
\nonumber\\
&\quad
-\frac{16\,\sigma_\rho^2\,\lambda_1\lambda_2(\rho_\infty-\rho_0)}{3(\tilde{\beta}^2+\Omega^2)^2}
\left[
-\frac{\tilde{A}_\beta}{2\tilde{\tau}^2}
+\frac{\tilde{B}_\beta\left(\frac{\tilde{\beta}^2}{2}-\frac{1}{2\tilde{\tau}^2}\right)}
{\frac{1}{\tilde{\tau}^2}+\Omega^2}
\right].
\end{align*}
If the initial local density is drawn from the stationary ensemble,
$\rho_0=\rho_\infty$, all terms proportional to
$(\rho_\infty-\rho_0)$ vanish identically. In this case,
$f$ drops out of Eq.\ref{eq:msd_short_time} and the early-time MSD reduces to
\begin{equation}
\langle \tilde r^2(t) \rangle
=
6\,\tilde t
+
\mathrm{Pe}^2\, g_{\mathrm{ss}}\,\tilde t^2 ,
\tag{29A}
\end{equation}
with
$g_{\mathrm{ss}}
=
(1-2\lambda_1\rho_\infty)
+
\sigma_\rho^2\,\lambda_1(\lambda_1-4\lambda_2\rho_\infty)$.
We now interpret the physical origin of the linear and quadratic
terms in Eq.\ref{eq:msd_short_time}.
 They encode contributions from memory effects and chirality-modulated orientation-density coupling. At very short times, the dynamics are dominated by the linear \( \tilde{t} \) term, representing passive thermal diffusion modified by transient local density gradients. The function \( f(\Omega, \tilde{\tau}, \sigma_\rho) \) captures how these gradients generate an effective drift that either enhances or suppresses diffusion depending on the sign of \( \rho_\infty - \rho_0 \). As time progresses, the quadratic \( \tilde{t}^2 \) term becomes significant, marking the onset of active ballistic motion. This contribution is governed by \( g(\Omega, \tilde{\tau}, \sigma_\rho) \), which incorporates the temporal memory of local density fluctuations, modulated by oscillatory orientation dynamics.

Importantly, the oscillatory signatures arising from chirality are strongest when the rotational and propulsion time scales are comparable—i.e., for intermediate values of \( \Omega \). In this regime, constructive interference in the orientation autocorrelations leads to a pronounced non-monotonic curvature in the MSD at early-to-intermediate times. For both low and high \( \Omega \), however, these oscillations are damped—either due to slow rotational decorrelation or rapid spinning—causing the short-time dynamics to appear more monotonic. This non-trivial dependence of \( g \) on \( \Omega \) reflects the interplay of persistence, chirality, and crowding in shaping the early-time active response.

In the opposite limit \( \tilde{t} \gg 1 \), the MSD asymptotes to a linear form \( \langle \tilde{\bm{r}}^2(\tilde{t}) \rangle \sim 6 D_{\mathrm{eff}} \tilde{t} \), where the effective dimensionless diffusivity is given by
\begin{equation}
\begin{aligned}
D_{\mathrm{eff}} =\; 1 
&+ \frac{\mathrm{Pe}^2}{18}(1 - 2\lambda_1 \rho_0)
\left[ 1 + \frac{8}{(4 + \Omega^2)} \right] \\
&+ \frac{2\mathrm{Pe}^2}{18} \sigma_\rho^2 \lambda_1(\lambda_1 - 4\lambda_2 \rho_0)
\left[ \frac{1}{\tilde{\beta}^2} + \frac{2}{(\tilde{\beta}^2 + \Omega^2)^2} \right],
\end{aligned}
\label{eq:msd_long_time}
\end{equation}
where \( \tilde{\beta} = 2 + \tilde{\tau}^{-1} \) is the dimensionless decay rate of density correlations.

The two additive terms beyond the thermal baseline \( D_{\mathrm{eff}} = 1 \) represent active contributions from persistent propulsion and from coupling to local density fluctuations. The first, deterministic term arises from propulsion in an interacting medium, modulated by chirality \( \Omega \) and local density \( \rho_0 \), while the second term encodes stochastic memory effects through density variance \( \sigma_\rho^2 \).

Two important physical limits are evident:

\textbf{(i) Interacting, achiral limit (\( \Omega \to 0 \)):}  
In this limit, the particle becomes non-chiral, but local steric interactions remain. The chirality-dependent terms simplify and yield an effective diffusivity:
\begin{align*}
D_{\mathrm{eff}} \to 1 + \frac{\mathrm{Pe}^2}{6}(1 - 2\lambda_1 \rho_0) 
+ \frac{\mathrm{Pe}^2}{3} \sigma_\rho^2 \lambda_1(\lambda_1 - 4\lambda_2 \rho_0) \cdot \frac{1}{\tilde{\beta}^2}.
\tag{30A}
\label{eq:diff_achiral}
\end{align*}

This highlights that even without chirality, interactions and memory lead to density-modulated diffusivity, with suppression in dense regions and amplification in dilute ones depending on the sign of the density coupling.

\textbf{(ii) Chiral, non-interacting limit (\( \rho_0 \to 0, \sigma_\rho^2 \to 0 \)):}  
Here, crowding and fluctuations vanish, and the expression reduces to
\begin{align*}
D_{\mathrm{eff}} \to 1 + \frac{\mathrm{Pe}^2}{18} \left(1 + \frac{8}{4 + \Omega^2} \right),
\tag{30B}
\label{eq:diff_chiral}
\end{align*}
For strong chirality \( \Omega \gg 1 \), the orientational memory is lost rapidly, and the active enhancement diminishes as \( D_{\mathrm{eff}} \sim 1 + \mathrm{Pe}^2/18 \). In contrast, for small chirality \( \Omega \ll 1 \), the enhanced persistence leads to the upper bound \( D_{\mathrm{eff}} \to 1 + \mathrm{Pe}^2/6 \), consistent with achiral free swimmers \cite{Howse2007}.

In general, chirality reduces long-time diffusivity by limiting persistent motion, while interactions introduce a non-monotonic dependence on local density and its fluctuations. These asymptotic forms fairly capture how propulsion, crowding, and angular dynamics conspire to determine the effective diffusion of chiral active particles in dense systems.
\begin{figure*}
    \begin{minipage}{0.98\linewidth} 
        \centering
        \includegraphics[width=\linewidth]{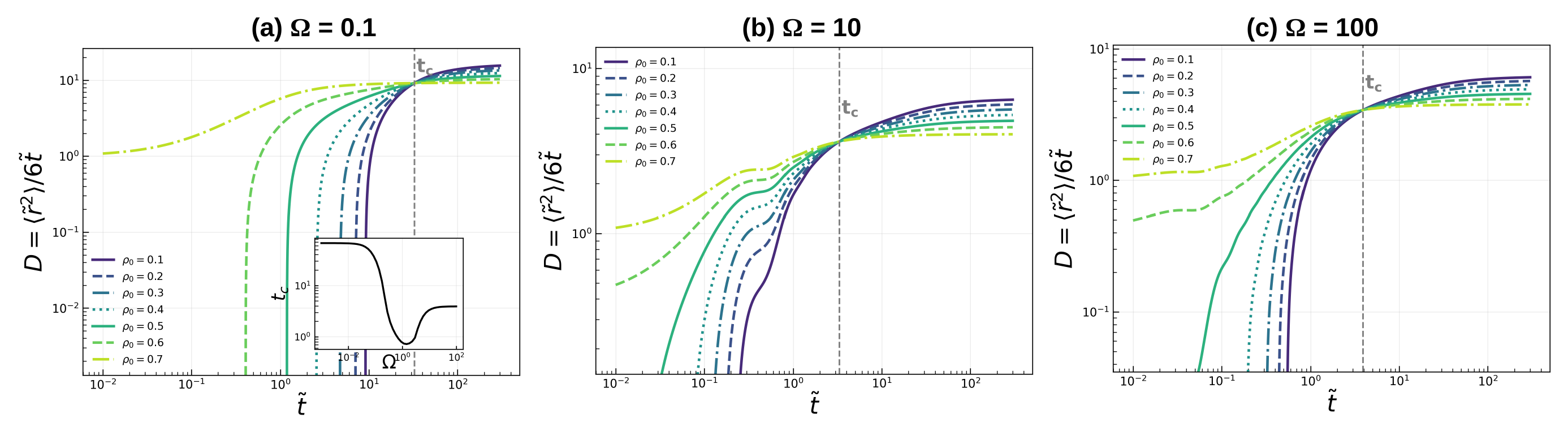}
        \caption{
    Time-dependent dimensionless diffusivity 
    \( D(t) = \langle \tilde{r}^2(t) \rangle / (6 \tilde{t}) \) 
    for various initial local densities \( \rho_0 \in [0.1, 0.7] \) at fixed global packing fraction 
    \( \rho_\infty = 0.70 \), translational diffusivity \( D_T = 0.1 \), rotational diffusivity \( D_r = 0.1 \), 
    self-propulsion speed \( v_0 = 1.0 \), fluctuation amplitude \( \sigma_\rho = 0.003 \), \(\tau = 100\) 
    and density coupling constants \( \lambda_1 = 1/(4\rho_\infty) \), \( \lambda_2 = 1/(16\rho_\infty^2) \). 
    (a), (b) and (c) show the analysis for three selected chiralities $\Omega$, highlighting the suppression of both diffusivities with increasing \( \Omega \). Vertical dashed lines mark the crossover time \( t_c \), defined as the minimum time after which the diffusivity \( D(t) \) for particles starting with initial local densities \( \rho_0 \) equalize. Inset: \( t_c \) versus \( \Omega \) (log–log axes) shows a non-monotonic dependence with a global minimum near \( \Omega \approx 1 \).}
    
        \label{fig:diffusivity-vs-time}
    \end{minipage}
\end{figure*}
\end{widetext}

\begin{figure}
    \centering
    \includegraphics[width=0.50\textwidth]{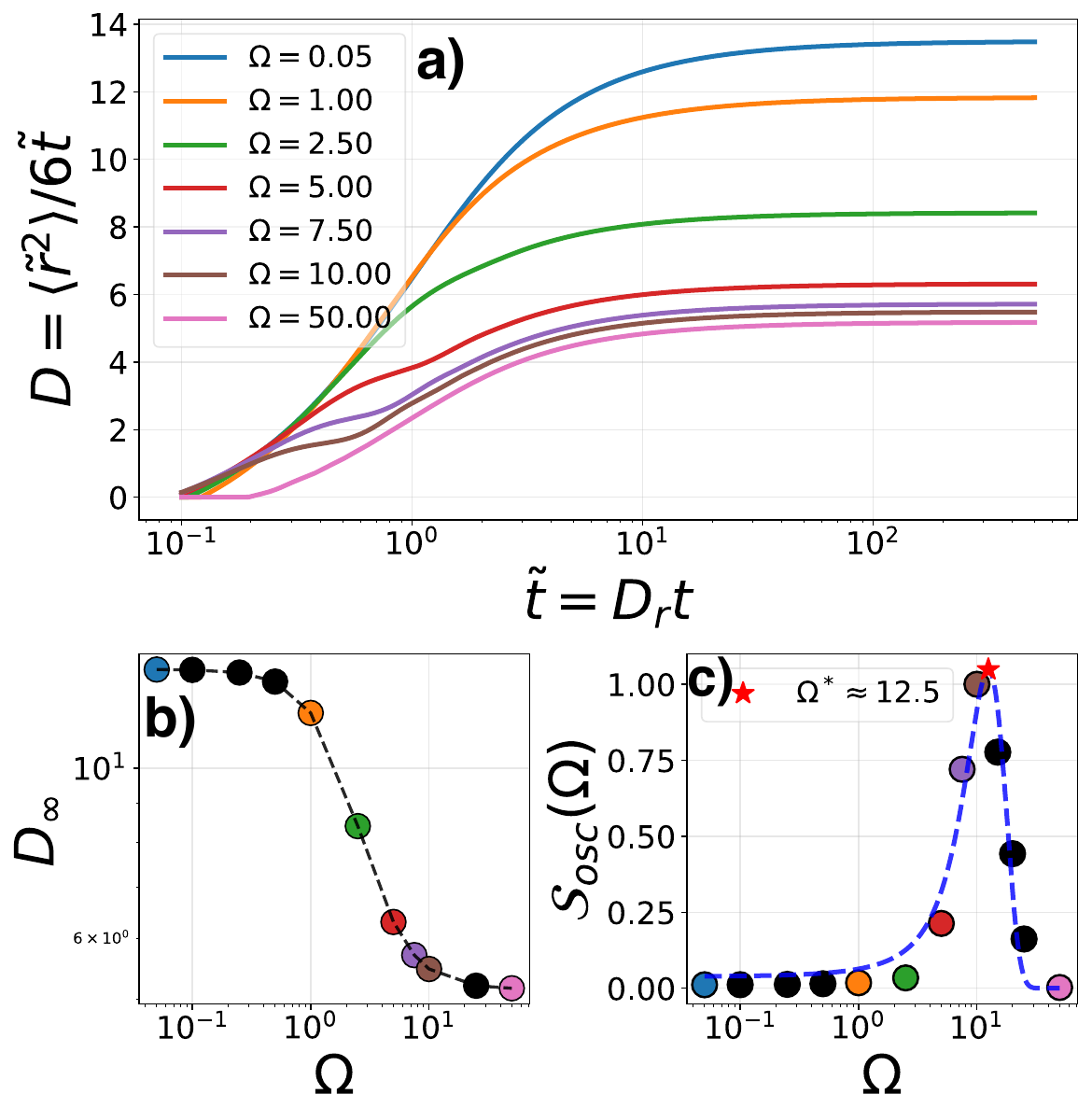}
    \caption{
(a) Time-dependent dimensionless diffusivity $D(\tilde t)=\langle\tilde r^2(\tilde t)\rangle/(6\tilde t)$ for selected chiralities $\Omega=\omega_0/D_r=\{0.05,1,2.5,5,7.5,10,50\}$. (b) Late-time diffusivity $D_\infty$ in the steady-state as a function of system chirality $\Omega$; axes are log–log to emphasize three distinct dynamics phases in chirality space. (c) Normalized oscillation strength $\mathcal S_{\rm osc}(\Omega)$ in the transient dynamics. $S_{osc}$ is calculated from the power spectral density of the oscillating part of the MSD. The dashed curve is a Gaussian fit used to extract the resonant chirality $\Omega^*$ (marker) at which maximum choerent rotational transport is achieved in the system. Parameters: $\rho_\infty=0.65$, initial local density averaged over all possible states $\langle \rho_0 \rangle_{\text{initial states}}  = \rho_\infty/2$, $D_r=0.1$, $D_T=0.1$, $v_0=1.0$ ($Pe=10$), $\tau=20.0$.}
    \label{fig:diffusivity-chirality}
\end{figure}

Figure~\ref{fig:diffusivity-vs-time} shows the analytically predicted time-dependent dimensionless diffusivity \( D(\tilde{t}) \) for particles starting in environments with different initial local densities \( \rho_0 \), while keeping all other system parameters fixed. For each chirality considered [panels (a--c)], the early-time dynamics depend sensitively on \( \rho_0 \). 
Notably, particles initialized in denser regions (\( \rho_0 > \rho_\infty/2 \)) display higher initial diffusivity than those in dilute regions. This anomalous enhancement arises from an active, non-equilibrium drift generated by gradients in the local swim pressure, set by the initial inhomogeneous density landscape. In the MSD expression, Eqn \ref{msd_opr_form}, these contributions are captured by the transient terms \( \mathcal{F}_\tau[\mathcal{C}_{\bm{\hat{n}}}](t) \) and \( \mathcal{F}_\tau[\mathcal{C}_{\bm{\hat{n}},\beta}](t) \), which encode the rapid rearrangement of density structures surrounding the particle. The curves for different \( \rho_0 \) converge at a well-defined crossover time \( t_c \), after which the the initial density inhomogeneity is largely erased and the system enters the relaxation stage toward the non-equilibrium steady state. Physically, \( t_c \) marks the timescale over which the excess active pressure generated by initial density gradients decays, removing the bias in short-time transport. For \( t > t_c \), the diffusivity curves—although starting from the same value at \( t_c \)—relax to different steady-state constants \( D_\infty(\rho_0) \). 
This residual separation indicates that the dynamics reflect a conditional dependence on the initial local environment, which does not persist when averaging over stationary initial conditions. 
In our model, this can be interpreted as arising from the fact that the long-time diffusivity depends not only on the instantaneous density field but also on the integrated history of how the particle's orientation decorrelated in the presence of past density fluctuations; in effect, the early-time environment influences the effective transport coefficients in the steady state. Repeating this analysis for a range of chiralities \( \Omega \) [panels (a--c)] shows that the overall diffusivity is suppressed at larger \( \Omega \), consistent with the reduced persistence length of chiral trajectories. 
The inset of panel (a) quantifies how \( t_c \) varies with chirality: it is large for nearly achiral systems, decreases sharply to a global minimum near \( \Omega \approx 1.0 \), and then saturates to a smaller constant value for highly chiral particles. 
This non-monotonic dependence suggests that intermediate chiralities induce transport in the particles that most efficiently randomize the local density field, breaking down the short-time density inhomogeneity in the environment faster than both achiral and strongly chiral limits.

Figure~\ref{fig:diffusivity-chirality} summarizes the effect of chirality on the dynamical response of the chiral active Brownian particle coupled to the fluctuating local density environment. Panel (a) shows the time-dependent dimensionless diffusivity $D(\tilde t)$ for a set of representative chiralities: all curves display the expected crossover from an early, oscillatory / ballistic-like transient into an asymptotic diffusive regime. Increasing $\Omega$ suppresses the long-time diffusivity, as chirality reduces orientational persistence and thereby shortens the effective run length; this suppression is strongest in the high-$\Omega$ limit where rotation dominates translational persistence.
\begin{figure}
\centering
\includegraphics[width=0.45\textwidth]{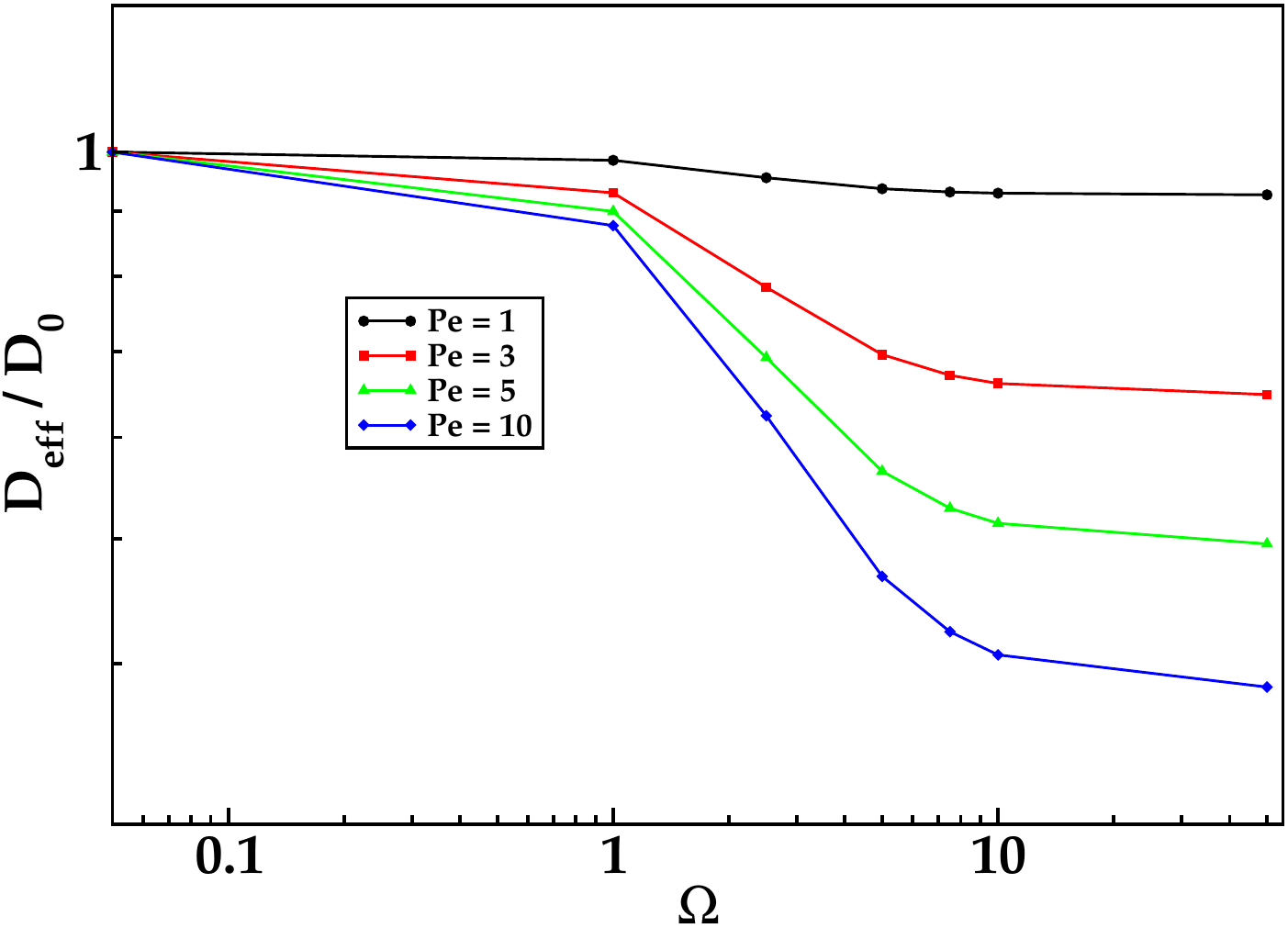}
\caption{Normalized steady-state diffusivity $D_{\mathrm{eff}}/D_0$ as a function of chirality $\Omega$ for different Péclet numbers $\mathrm{Pe} = 1, 3, 5, 10$, obtained from the analytical theory. Here $D_0$ corresponds to the case of the particle being non chiral or $\Omega = 0$,derived from Eq.\ref{eq:diff_achiral}}
\label{fig:diff_omega.pdf}
\end{figure}
Panel (b) quantifies this suppression by plotting the late-time diffusivity $D_\infty$ as a function of chirality $\Omega$ in the system. Three regimes are apparent: (i) for very small $\Omega$ the diffusivity is approximately constant (weak chirality limit), (ii) for intermediate $\Omega$ the curve falls rapidly and forms a broad plateau-like region consistent with an emergent micro-clustered dynamical phase, and (iii) for very large $\Omega$ transport is strongly suppressed and $D_\infty\to0$. The plateau in regime (ii) agrees qualitatively with earlier observations of micro-clustering in intermediate-chirality swimmers \cite{Semwal2024}. To further benchmark our analytical predictions against earlier numerical studies, we compare the chirality dependence of the steady-state diffusivity obtained from the present theory with results from our previous many-particle simulations. Figure \ref{fig:diff_omega.pdf} shows the normalized long-time diffusivity $D_{\mathrm{eff}}/D_0$ as a function of chirality $\Omega$ for several Péclet numbers. The theory exhibits qualitative trends consistent with those reported in \cite{Semwal2024}. This comparison indicates that the present analytical framework captures qualitatively similar chirality-dependent transport behavior, providing a consistency check that the density–coupled single-particle description reflects key features observed in many-body simulations.

Panel~(c) quantifies the observed oscillatory signature in the particle dynamics in the transient-time regime for different values of chirality. We measure the strength of oscillatory dynamics which provides insight about the extent of cohorent rotational transport in the dynamics  $\mathcal{S}_{\mathrm{osc}}(\Omega)$, by first isolating the oscillatory signal obtained by subtracting the smooth ballistic trend from the MSD. Then, computing the power spectral density (PSD) of the residual, and integrating over the oscillatory frequency band we obtain $\mathcal{S}_{osc} = \int PSD\big[\langle \tilde{r}_{osc}^2\rangle\big](\nu) d\nu$. $\mathcal{S}_{\mathrm{osc}}$ peaks at a resonant $\Omega^*$, indicating that coherent oscillations are maximal when the active rotation rate matches the effective decorrelation rate $\beta = 2D_r + 1/\tau$.  
For $\Omega \ll \Omega^*$, density-memory damping suppresses oscillations before a rotation is completed, while for $\Omega \gg \Omega^*$ rapid reorientation disrupts their buildup.

\begin{figure}[t]
    \centering
    \includegraphics[width=0.95\linewidth]{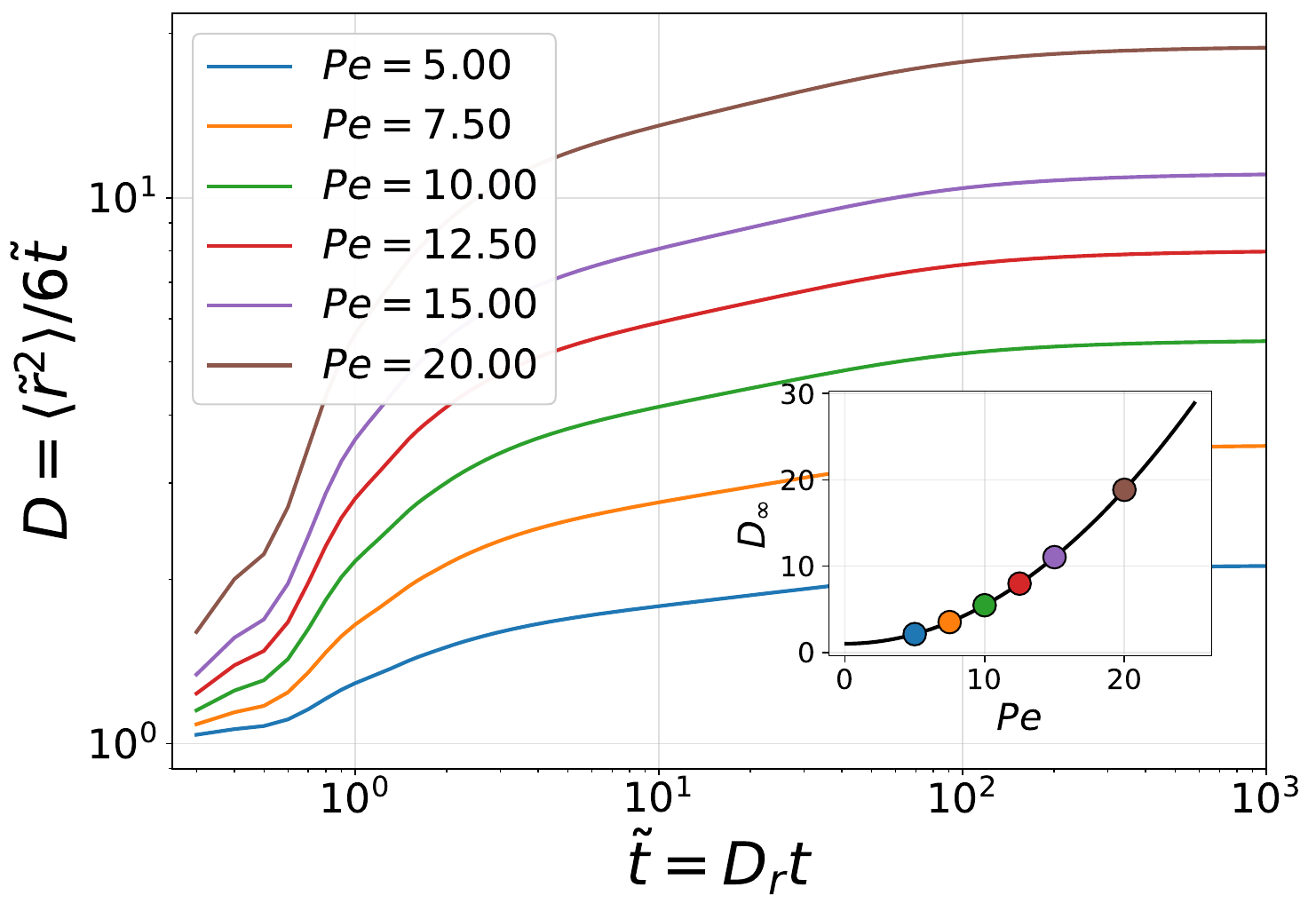}
    \caption{
    (Main) Time-dependent dimensionless diffusivity 
    \( D(t) = \langle \tilde{r}^2(t) \rangle / (6 \tilde{t}) \) 
    for varying Péclet numbers \( Pe = v_0 / \sqrt{D_T D_R} \) at fixed chirality 
    \( \Omega = \omega_0 / D_R = 5.0 \), global density \( \rho_\infty = 0.70 \), 
    and initial density \( \rho_0 = \rho_\infty/2 \).  
    Increasing activity systematically enhances diffusivity across all timescales, 
    with the largest relative change observed in the steady-state regime.  
    (Inset) Late-time diffusivity \( D_\infty \), computed as the average of \( D(t) \) over the final 5\% of simulation time, 
    as a function of \( Pe \). The inset shows a monotonic growth of \( D_\infty \) with activity, 
    with the slope decreasing at large \( Pe \), indicating a saturation of activity-induced transport enhancement.
    Model parameters: \( D_R = 0.1 \), \( D_T = 0.1 \), \( D_R \tau = 20.0 \), \( \sigma_\rho = 0.1 \), 
    \( \lambda_1 = 1/(4\rho_\infty) \), \( \lambda_2 = 1/(16\rho_\infty^2) \).
    }
    \label{fig:diffusivity-activity}
\end{figure}
Figure~\ref{fig:diffusivity-activity} presents the time-dependent dimensionless diffusivity 
$D(t) = \langle \tilde{r}^2(t) \rangle / (6 \tilde{t})$ for a range of Péclet numbers 
$Pe = v_0 / \sqrt{D_T D_R}$, at fixed chirality $\Omega = 10.0$ and density parameters 
$(\rho_\infty, \rho_0) = (0.70, \rho_\infty / 2)$. 
The trends are consistent with those reported for dense suspensions of chiral ABPs: 
increasing activity enhances both the magnitude of the early-time oscillatoy ballistic regime and 
the late-time steady-state diffusivity $D_\infty$. 
The inset quantifies this latter effect by plotting $D_\infty$ as a function of $Pe$; 
the monotonic growth $(\propto Pe^2)$ confirms that persistent propulsion directly promotes transport in the steady-state. While the observed $D(t)$ and $D_\infty(Pe)$ behaviours are not themselves qualitatively novel, their agreement with established numerical and experimental results \cite{aparna2008prl} for chiral active particles provides an important consistency check for our analytical framework. In particular, it supports our modelling strategy in which steric particle interactions are coarse-grained into an effective stochastic local density field with correlation time $\tau$ and variance $\sigma_\rho^2$. \\

\section{Discussion}

We have developed a closed-form analytical theory for the transport of a chiral active Brownian particle (cABP) in a fluctuating local density field with finite correlation time $\tau$ and fluctuation strength $\sigma_\rho^2$. This coarse-grained modelling replaces explicit inter-particle steric interactions in a dense active medium with an Ornstein--Uhlenbeck process for the local density, fluctuating about a mean value and relaxing towards a global steady state. This construction captures both the transient dynamics preceding the non-equilibrium steady state and the persistent statistical structure of the environment, while allowing fully tractable analytical expressions for the MSD and time-dependent diffusivity.

Our analytical result of MSD is qualitatively consistent with known trends observed in simulations and experiments, supporting the plausibility of the framework. Within this setting, we investigate how the particle's orientational diffusion couples to local density fluctuations, and how chirality modifies these dynamics in nontrivial ways.

First, analysing the early- and long-time behaviour of the diffusivity reveals a clear dependence on the initial local density $\rho_0$. Particles starting in denser regions exhibit anomalously high initial diffusivity due to transient active drift generated by local swim-pressure gradients, as encoded in the transient terms of the MSD expression. We identify a finite crossover time $t_c$ at which these drifts homogenise the initial density inhomogeneities. The dependence on $\rho_0$ should be interpreted as a conditional effect associated with the specified initial local environment. When the initial condition is drawn from the stationary ensemble ($\rho_0 = \rho_\infty$), this dependence vanishes, and the steady-state diffusivity becomes independent of the initial density basin. Thus, the effect does not represent a generic steady-state memory, but rather reflects transient dynamics conditioned on the initial state.

Second, varying chirality reveals a non-monotonic $t_c(\Omega)$ with a global minimum at intermediate $\Omega$, corresponding to an optimal chirality that most efficiently erases initial density contrasts. The long-time diffusivity $D_\infty(\Omega)$ displays three distinct regimes: a low-$\Omega$ plateau, a gradual suppression over intermediate $\Omega$ values, and a high-$\Omega$ plateau approaching zero diffusivity. The intermediate-$\Omega$ suppression regime shows qualitative consistency with micro-clustered phases previously reported in simulations of chiral active swimmers~\cite{Semwal2024}, providing further agreement with the coarse-grained approach.

Finally, by isolating the oscillatory component of the early-time MSD, we quantify the oscillation strength $S_{\mathrm{osc}}(\Omega)$ and find a resonance-like peak at $\Omega^*$. This optimal chirality results from the interplay between orientational diffusion, density-field decorrelation, and the imposed circular motion, producing maximally coherent rotational transport at early times. Such a peak may be relevant for understanding the emergence of optimal chirality in dense colonies of biological circle swimmers, and could provide qualitative insight into how transport efficiency might be influenced by rotational dynamics.

Beyond recovering established trends for $D_\infty$ versus $Pe$ and $\Omega$, our theory uncovers new dynamical signatures, including a conditional dependence on the initial local environment and the existence of an optimal chirality for coherent oscillations, all within a minimal, analytically tractable framework. This approach offers both a compact physical interpretation of simulation and experimental data, and a framework that can be useful for exploring transport in biological chiral active matter as well as designing active meta-materials.

The present theory is formulated in the overdamped limit, which is appropriate for microscale active systems where inertial effects are negligible. In this regime, particle velocities are instantaneously slaved to the propulsion direction, allowing a factorization of velocity correlations into independent speed and orientation contributions. An extension of the present framework to include inertia would require incorporating underdamped dynamics, leading to additional velocity relaxation timescales and modifying the structure of the velocity autocorrelation function. 

\section{Conflicts of Interest}
There are no conflicts of interest to declare.
\section{Acknowledgement}
The support and the resources provided by PARAM
Shivay Facility under the National Supercomputing Mision, Government of India at the Indian Institute of
Technology, Varanasi are gratefully acknowledged by all authors. SM thanks DSTSERB India, ECR/2017/000659, CRG/2021/006945 and
MTR/2021/000438 for financial support.
\appendix

\section{Derivation of the Angular Equations in It\^o Calculus}\label{app:A}
This appendix derives the stochastic evolution equations for the spherical angles
$\theta(t)$ and $\phi(t)$ of the unit orientation vector $\hat{\mathbf n}(t)$,
starting from the vector Langevin equation supplied in the main text. 

\subsection{Vector equation and its It\^o form}

We start from the stochastic orientation dynamics written in the form
\begin{equation}
d\hat{\mathbf n}
=
\big(\hat{\mathbf n}\times \omega_0 \hat{\mathbf z}\big)\,dt
+
\sqrt{2D_r}\,\big(\hat{\mathbf n}\times d\mathbf W\big),
\label{eq:A_start}
\end{equation}
where $d\mathbf W=(dW_x,dW_y,dW_z)$ is a three-dimensional Wiener increment with
\begin{equation}
\langle dW_i\,dW_j\rangle=\delta_{ij}\,dt,
\label{eq:A_Wiener}
\end{equation}
and the It\^o multiplication rules are
\begin{equation}
(dt)^2=0,\qquad dt\,dW_i=0,\qquad dW_i\,dW_j=\delta_{ij}\,dt.
\label{eq:A_ItoRules}
\end{equation}

To preserve the unit-length constraint $\hat{\mathbf n}\cdot\hat{\mathbf n}=1$ in the
It\^o interpretation, we impose $d(\hat{\mathbf n}\cdot\hat{\mathbf n})=0$.
Using the It\^o product rule,
\begin{equation}
d(\hat{\mathbf n}\cdot\hat{\mathbf n})
=
2\,\hat{\mathbf n}\cdot d\hat{\mathbf n}
+
d\hat{\mathbf n}\cdot d\hat{\mathbf n}.
\label{eq:A_n2}
\end{equation}
If Eq.~\eqref{eq:A_start} is treated naively as an It\^o SDE, then
$\hat{\mathbf n}\cdot(\hat{\mathbf n}\times\hat{\mathbf z})=0$ and
$\hat{\mathbf n}\cdot(\hat{\mathbf n}\times d\mathbf W)=0$, so that
$2\hat{\mathbf n}\cdot d\hat{\mathbf n}=0$.
However, the quadratic term in \eqref{eq:A_n2} is nonzero: keeping only noise--noise
products (since $(dt)^2=dt\,dW_i=0$),
\begin{equation}
d\hat{\mathbf n}\cdot d\hat{\mathbf n}
=
2D_r\,|\hat{\mathbf n}\times d\mathbf W|^2.
\label{eq:A_dndn_1}
\end{equation}
Using $|\mathbf a\times \mathbf b|^2=|\mathbf a|^2|\mathbf b|^2-(\mathbf a\cdot\mathbf b)^2$
with $|\hat{\mathbf n}|=1$,
\begin{equation}
|\hat{\mathbf n}\times d\mathbf W|^2
=
|d\mathbf W|^2-(\hat{\mathbf n}\cdot d\mathbf W)^2.
\label{eq:A_cross_identity}
\end{equation}
From \eqref{eq:A_ItoRules},
\begin{equation}
|d\mathbf W|^2=dW_x^2+dW_y^2+dW_z^2=3\,dt,
\label{eq:A_dWsq}
\end{equation}
and
\begin{align}
(\hat{\mathbf n}\cdot d\mathbf W)^2
&=(n_x dW_x+n_y dW_y+n_z dW_z)^2 \nonumber\\
&=n_x^2 dW_x^2+n_y^2 dW_y^2+n_z^2 dW_z^2
\nonumber\\
&\quad
+2n_xn_y\,dW_x dW_y+2n_xn_z\,dW_x dW_z
\nonumber\\
&\quad
+2n_yn_z\,dW_y dW_z \nonumber\\
&=(n_x^2+n_y^2+n_z^2)\,dt
=dt,
\label{eq:A_proj_sq}
\end{align}
where the mixed terms vanish because $dW_i dW_j=0$ for $i\neq j$ in It\^o calculus and
$n_x^2+n_y^2+n_z^2=1$.
Equations \eqref{eq:A_cross_identity}--\eqref{eq:A_proj_sq} give
$|\hat{\mathbf n}\times d\mathbf W|^2=2dt$, hence
\begin{equation}
d\hat{\mathbf n}\cdot d\hat{\mathbf n}=4D_r\,dt,
\qquad\Rightarrow\qquad
d(\hat{\mathbf n}\cdot\hat{\mathbf n})=4D_r\,dt\neq 0,
\end{equation}
showing that \eqref{eq:A_start} cannot be the correct It\^o form.

To restore $d(\hat{\mathbf n}\cdot\hat{\mathbf n})=0$, an additional drift term is required.
The rotationally invariant choice is proportional to $\hat{\mathbf n}$; we write it as
$-\lambda \hat{\mathbf n}\,dt$ and determine $\lambda$.
With
\begin{equation}
d\hat{\mathbf n}
=
\big(\hat{\mathbf n}\times \omega_0 \hat{\mathbf z}-\lambda \hat{\mathbf n}\big)\,dt
+
\sqrt{2D_r}\,\big(\hat{\mathbf n}\times d\mathbf W\big),
\label{eq:A_trialIto}
\end{equation}
one has $2\hat{\mathbf n}\cdot d\hat{\mathbf n}=-2\lambda\,dt$, while the quadratic term
remains $d\hat{\mathbf n}\cdot d\hat{\mathbf n}=4D_r\,dt$ as above. Substituting in
\eqref{eq:A_n2} and imposing $d(\hat{\mathbf n}\cdot\hat{\mathbf n})=0$ yields
$-2\lambda+4D_r=0$, hence $\lambda=2D_r$.
Therefore, the It\^o equation equivalent to \eqref{eq:A_start} and consistent with
$|\hat{\mathbf n}|=1$ is
\begin{equation}
\boxed{
d\hat{\mathbf n}
=
\big(\hat{\mathbf n}\times \omega_0 \hat{\mathbf z}-2D_r\,\hat{\mathbf n}\big)\,dt
+
\sqrt{2D_r}\,\big(\hat{\mathbf n}\times d\mathbf W\big).
}
\label{eq:A_ItoVectorFinal}
\end{equation}

\subsection{Spherical parametrization}

We parameterize the unit vector by spherical angles,
\begin{equation}
\hat{\mathbf n}
=
(\sin\theta\cos\phi,\;\sin\theta\sin\phi,\;\cos\theta),
\label{eq:A_spherical}
\end{equation}
so that
\begin{equation}
n_x=\sin\theta\cos\phi,\qquad
n_y=\sin\theta\sin\phi,\qquad
n_z=\cos\theta.
\label{eq:A_components}
\end{equation}
We now derive the It\^o SDEs for $\theta$ and $\phi$ directly from
Eq.~\eqref{eq:A_ItoVectorFinal}.

\subsection{Derivation of $d\theta$}

Since $n_z=\cos\theta$, equivalently $\theta=\cos^{-1}(n_z)$, we apply It\^o's lemma for a
function of a single stochastic variable:
\begin{equation}
d\theta
=
\frac{d\theta}{dn_z}\,dn_z
+
\frac{1}{2}\frac{d^2\theta}{dn_z^2}\,(dn_z)^2.
\label{eq:A_ItoTheta}
\end{equation}
From $\theta=\cos^{-1}(n_z)$,
\begin{equation}
\frac{d\theta}{dn_z}=-\frac{1}{\sqrt{1-n_z^2}},
\qquad
\frac{d^2\theta}{dn_z^2}=-\frac{n_z}{(1-n_z^2)^{3/2}}.
\label{eq:A_theta_derivs_nz}
\end{equation}
Using $n_z=\cos\theta$ and $1-n_z^2=\sin^2\theta$,
\begin{equation}
\frac{d\theta}{dn_z}=-\frac{1}{\sin\theta},
\qquad
\frac{d^2\theta}{dn_z^2}=-\frac{\cos\theta}{\sin^3\theta}.
\label{eq:A_theta_derivs_theta}
\end{equation}

The $z$-component of Eq.~\eqref{eq:A_ItoVectorFinal} is obtained by noting that
$(\hat{\mathbf n}\times\hat{\mathbf z})_z=0$ and
$(\hat{\mathbf n}\times d\mathbf W)_z=n_x dW_y-n_y dW_x$, giving
\begin{equation}
dn_z
=
-2D_r n_z\,dt+\sqrt{2D_r}\,(n_x dW_y-n_y dW_x).
\label{eq:A_dnz}
\end{equation}
To evaluate $(dn_z)^2$, we retain only the noise--noise contribution:
\begin{align}
(dn_z)^2
&=
2D_r\,(n_x dW_y-n_y dW_x)^2 \nonumber\\
&=
2D_r\Big(n_x^2 dW_y^2+n_y^2 dW_x^2-2n_x n_y\,dW_y dW_x\Big) \nonumber\\
&=
2D_r\,(n_x^2+n_y^2)\,dt
=
2D_r\sin^2\theta\,dt,
\label{eq:A_dnz_sq}
\end{align}
where $dW_x^2=dW_y^2=dt$ and $dW_y dW_x=0$ were used, together with
$n_x^2+n_y^2=\sin^2\theta$ from \eqref{eq:A_components}.

Substituting \eqref{eq:A_theta_derivs_theta}, \eqref{eq:A_dnz}, and \eqref{eq:A_dnz_sq} into
\eqref{eq:A_ItoTheta} yields
\begin{align}
d\theta
&=
\left(-\frac{1}{\sin\theta}\right)\left[-2D_r\cos\theta\,dt+\sqrt{2D_r}(n_x dW_y-n_y dW_x)\right]
\nonumber\\
&\quad
+\frac{1}{2}\left(-\frac{\cos\theta}{\sin^3\theta}\right)\left(2D_r\sin^2\theta\,dt\right)
\nonumber\\
&=
\Big(2D_r\cot\theta-D_r\cot\theta\Big)\,dt
-\frac{\sqrt{2D_r}}{\sin\theta}\,(n_x dW_y-n_y dW_x).
\label{eq:A_dtheta_expand}
\end{align}
Thus the drift simplifies to $D_r\cot\theta\,dt$.
The remaining stochastic term is a linear combination of $dW_x$ and $dW_y$; it can be written
as a single Wiener increment $dW_\theta$ defined by
\begin{equation}
dW_\theta
\equiv
-\frac{1}{\sin\theta}(n_x dW_y-n_y dW_x),
\qquad
\langle dW_\theta^2\rangle=dt.
\label{eq:A_dWtheta_def}
\end{equation}
We therefore obtain the It\^o equation for $\theta$:
\begin{equation}
\boxed{
d\theta
=
D_r\cot\theta\,dt+\sqrt{2D_r}\,dW_\theta.
}
\label{eq:A_theta_final}
\end{equation}

\subsection{Derivation of $d\phi$}

The azimuthal angle is defined by
\begin{equation}
\phi=\tan^{-1}\!\left(\frac{n_y}{n_x}\right),
\label{eq:A_phi_def}
\end{equation}
and is therefore a function of the two stochastic variables $(n_x,n_y)$.
We apply It\^o's lemma in two variables:
\begin{align}
d\phi
&=
\frac{\partial\phi}{\partial n_x}\,dn_x
+\frac{\partial\phi}{\partial n_y}\,dn_y
\nonumber\\
&
+\frac{1}{2}\left[
\frac{\partial^2\phi}{\partial n_x^2}(dn_x)^2
+2\frac{\partial^2\phi}{\partial n_x\partial n_y}\,dn_x dn_y
+\frac{\partial^2\phi}{\partial n_y^2}(dn_y)^2
\right].
\label{eq:A_ItoPhi}
\end{align}

Introduce $R^2\equiv n_x^2+n_y^2$. Differentiating \eqref{eq:A_phi_def} gives the first derivatives
\begin{equation}
\frac{\partial\phi}{\partial n_x}=-\frac{n_y}{R^2},
\qquad
\frac{\partial\phi}{\partial n_y}=\frac{n_x}{R^2},
\label{eq:A_phi_first}
\end{equation}
and the second derivatives (kept explicit here, as they enter the It\^o correction term) are
\begin{align}
&\frac{\partial^2\phi}{\partial n_x^2}=\frac{2n_x n_y}{R^4},
\qquad
\frac{\partial^2\phi}{\partial n_y^2}=-\frac{2n_x n_y}{R^4},
\nonumber\\
&
\frac{\partial^2\phi}{\partial n_x\partial n_y}=\frac{n_y^2-n_x^2}{R^4}.
\label{eq:A_phi_second}
\end{align}

Next we extract $dn_x$ and $dn_y$ from the vector It\^o equation \eqref{eq:A_ItoVectorFinal}.
Using $\hat{\mathbf n}\times\hat{\mathbf z}=(n_y,-n_x,0)$ and
\begin{align}
&(\hat{\mathbf n}\times d\mathbf W)_x=n_y dW_z-n_z dW_y
\nonumber\\
&
(\hat{\mathbf n}\times d\mathbf W)_y=n_z dW_x-n_x dW_z,
\nonumber
\end{align}
the $x$ and $y$ components are
\begin{equation}
dn_x=(\omega_0 n_y-2D_r n_x)\,dt+\sqrt{2D_r}\,(n_y dW_z-n_z dW_y),
\label{eq:A_dnx}
\end{equation}
\begin{equation}
dn_y=(-\omega_0 n_x-2D_r n_y)\,dt+\sqrt{2D_r}\,(n_z dW_x-n_x dW_z).
\label{eq:A_dny}
\end{equation}

We first compute the drift contribution from the first-derivative terms in
Eq.~\eqref{eq:A_ItoPhi}. Substituting the $dt$ parts of \eqref{eq:A_dnx}--\eqref{eq:A_dny}
into \eqref{eq:A_phi_first} gives
\begin{align}
d\phi_{\mathrm{drift}}
&=
\left(-\frac{n_y}{R^2}\right)(\omega_0 n_y-2D_r n_x)\,dt
\nonumber\\
&\quad
+\left(\frac{n_x}{R^2}\right)(-\omega_0 n_x-2D_r n_y)\,dt \nonumber\\
&=
-\omega_0\frac{n_x^2+n_y^2}{R^2}\,dt
+\frac{2D_r n_x n_y-2D_r n_x n_y}{R^2}\,dt
\nonumber\\
&
=
-\omega_0\,dt.
\label{eq:A_phi_drift}
\end{align}
Thus the deterministic angular velocity is $-\omega_0$ (consistent with the convention
$\partial_t\hat{\mathbf n}=\hat{\mathbf n}\times\omega_0\hat{\mathbf z}$ used in the main text).

The remaining task is to evaluate the second-order It\^o correction and the noise term.
For the quadratic variations we retain only noise--noise contributions from
\eqref{eq:A_dnx}--\eqref{eq:A_dny}. Using \eqref{eq:A_ItoRules} one finds
\begin{align}
(dn_x)^2
&=
2D_r\,(n_y dW_z-n_z dW_y)^2
=2D_r\,(n_y^2+n_z^2)\,dt,
\label{eq:A_dnx2}
\\
(dn_y)^2
&=
2D_r\,(n_z dW_x-n_x dW_z)^2
=2D_r\,(n_z^2+n_x^2)\,dt,
\label{eq:A_dny2}
\\
dn_x\,dn_y
&=
2D_r\,(n_y dW_z-n_z dW_y)(n_z dW_x-n_x dW_z)
\nonumber\\
&=
-2D_r\,n_x n_y\,dt,
\label{eq:A_dnx_dny}
\end{align}
where the only surviving products are $dW_i^2=dt$.

Substituting \eqref{eq:A_phi_second} and \eqref{eq:A_dnx2}--\eqref{eq:A_dnx_dny} into the
second-order bracket in \eqref{eq:A_ItoPhi}, we obtain
\begin{align}
&\frac{1}{2}\Big[
\phi_{xx}(dn_x)^2+2\phi_{xy}dn_x dn_y+\phi_{yy}(dn_y)^2
\Big]
\nonumber\\
&=
\frac{1}{2}\frac{1}{R^4}
\Big[
(2n_x n_y)\,2D_r(n_y^2+n_z^2)\,dt
\nonumber\\
&
+2(n_y^2-n_x^2)\,(-2D_r n_x n_y\,dt)
+(-2n_x n_y)\,2D_r(n_z^2+n_x^2)\,dt
\Big]
\nonumber\\
&=
\frac{2D_r n_x n_y}{R^4}
\Big[(n_y^2+n_z^2)-(n_y^2-n_x^2)-(n_z^2+n_x^2)\Big]\,dt
=0.
\label{eq:A_phi_Ito_cancel}
\end{align}
Hence the It\^o second-order correction contributes no additional drift to $\phi$.

Finally, the stochastic part of $d\phi$ arises from the first-derivative terms in
\eqref{eq:A_ItoPhi}. Substituting the noise parts of \eqref{eq:A_dnx}--\eqref{eq:A_dny}
into \eqref{eq:A_phi_first} yields
\begin{align}
d\phi_{\mathrm{noise}}
&=
-\frac{n_y}{R^2}\sqrt{2D_r}\,(n_y dW_z-n_z dW_y)
\nonumber\\
&
+\frac{n_x}{R^2}\sqrt{2D_r}\,(n_z dW_x-n_x dW_z)
\nonumber\\
&=
\frac{\sqrt{2D_r}}{R^2}\Big(n_x n_z\,dW_x+n_y n_z\,dW_y-(n_x^2+n_y^2)\,dW_z\Big).
\label{eq:A_phi_noise_raw}
\end{align}
Since $R^2=n_x^2+n_y^2=\sin^2\theta$, this can be written in the standard form
\begin{equation}
d\phi_{\mathrm{noise}}
=
\frac{\sqrt{2D_r}}{\sin\theta}\,dW_\phi,
\qquad
\langle dW_\phi^2\rangle=dt,
\label{eq:A_phi_noise_def}
\end{equation}
where $dW_\phi$ is a one-dimensional Wiener increment obtained as the appropriate linear
combination of $(dW_x,dW_y,dW_z)$ in \eqref{eq:A_phi_noise_raw}. (A change of sign
$dW_\phi\to -dW_\phi$ is immaterial.)

Combining \eqref{eq:A_phi_drift}, \eqref{eq:A_phi_Ito_cancel}, and \eqref{eq:A_phi_noise_def}
gives the It\^o equation for $\phi$:
\begin{equation}
\boxed{
d\phi
=
-\omega_0\,dt+\frac{\sqrt{2D_r}}{\sin\theta}\,dW_\phi.
}
\label{eq:A_phi_final}
\end{equation}

\subsection{Final angular equations}

Collecting Eqs.~\eqref{eq:A_theta_final} and \eqref{eq:A_phi_final}, the It\^o angular dynamics read
\begin{equation}
\boxed{
\begin{aligned}
d\theta &= D_r\cot\theta\,dt + \sqrt{2D_r}\,dW_\theta,\\[4pt]
d\phi &= -\omega_0\,dt + \frac{\sqrt{2D_r}}{\sin\theta}\,dW_\phi,
\end{aligned}}
\qquad
\langle dW_\alpha dW_\beta\rangle=\delta_{\alpha\beta}\,dt.
\end{equation}
These equations describe rotational diffusion of $\hat{\mathbf n}$ on the unit sphere with a
constant chiral rotation about the $z$ axis.

\section{Rotation operator on the unit sphere}
\label{app:B}


In the main text we use the rotation operator
\begin{equation}
\bm R \equiv \hat{\bm n}\times \nabla_{\hat{\bm n}},
\label{eq:Rdef_app}
\end{equation}
which generates infinitesimal rotations of the unit orientation vector
\(\hat{\bm n}\). Here \(\nabla_{\hat{\bm n}}\) denotes the angular gradient
on the unit sphere.

\subsection{Angular coordinates and surface gradient}

We parametrize the unit vector by spherical angles as
\begin{equation}
\hat{\bm n}(\theta,\phi)=
(\sin\theta\cos\phi)\,\hat{\bm x}
+(\sin\theta\sin\phi)\,\hat{\bm y}
+(\cos\theta)\,\hat{\bm z}.
\label{eq:n_spherical_app}
\end{equation}

The tangent basis vectors on the unit sphere are
\begin{equation}
\partial_\theta \hat{\bm n} \equiv \hat{\bm e}_\theta,
\qquad
\partial_\phi \hat{\bm n} = \sin\theta\,\hat{\bm e}_\phi,
\label{eq:tangent_basis_app}
\end{equation}
where \(\hat{\bm e}_\theta\) and \(\hat{\bm e}_\phi\) are orthonormal and
tangent to the sphere.

For any scalar function \(f(\theta,\phi)\) defined on the unit sphere,
the angular (surface) gradient is
\begin{equation}
\nabla_{\hat{\bm n}} f
=
\hat{\bm e}_\theta\,\partial_\theta f
+
\hat{\bm e}_\phi\,\frac{1}{\sin\theta}\,\partial_\phi f.
\label{eq:angular_gradient_app}
\end{equation}

The basis vectors satisfy the identities
\begin{equation}
\hat{\bm n}\times \hat{\bm e}_\theta = \hat{\bm e}_\phi,
\qquad
\hat{\bm n}\times \hat{\bm e}_\phi = -\hat{\bm e}_\theta.
\label{eq:cross_identities_app}
\end{equation}

\subsection{Action of the rotation operator}

Using Eqs.~\eqref{eq:Rdef_app} and \eqref{eq:angular_gradient_app},
the action of \(\bm R\) on a scalar function \(f(\theta,\phi)\) is
\begin{align}
\bm R f
&= \hat{\bm n}\times (\nabla_{\hat{\bm n}} f) \nonumber\\
&= \hat{\bm n}\times
\left(
\hat{\bm e}_\theta\,\partial_\theta f
+
\hat{\bm e}_\phi\,\frac{1}{\sin\theta}\partial_\phi f
\right) \nonumber\\
&= \hat{\bm e}_\phi\,\partial_\theta f
-
\hat{\bm e}_\theta\,\frac{1}{\sin\theta}\partial_\phi f,
\label{eq:R_action_app}
\end{align}
where Eq.~\eqref{eq:cross_identities_app} has been used.

\subsection{Explicit coordinate form of \(R^2\)}

We define
\begin{equation}
R^2 \equiv \bm R\cdot \bm R.
\label{eq:R2def_app}
\end{equation}
The operator \(R^2\) coincides with the Laplace--Beltrami operator on the
unit sphere, yielding the explicit coordinate representation
\begin{equation}
R^2 f(\theta,\phi)
=
\frac{1}{\sin\theta}\,\partial_\theta
\left(
\sin\theta\,\partial_\theta f
\right)
+
\frac{1}{\sin^2\theta}\,\partial_\phi^2 f.
\label{eq:R2_coord_app}
\end{equation}

\subsection{Proof of \(R^2\hat{\bm n}=-2\hat{\bm n}\)}

Using Eq.~\eqref{eq:n_spherical_app}, the components of
\(\hat{\bm n}\) are
\[
n_x=\sin\theta\cos\phi,\qquad
n_y=\sin\theta\sin\phi,\qquad
n_z=\cos\theta.
\]

\paragraph{(i) \(n_z=\cos\theta\).}
Since \(\partial_\theta n_z=-\sin\theta\) and
\(\partial_\phi n_z=0\),
\begin{align}
R^2(\cos\theta)
&=
\frac{1}{\sin\theta}\partial_\theta
\left(
\sin\theta\,\partial_\theta\cos\theta
\right) \nonumber\\
&=
\frac{1}{\sin\theta}\partial_\theta(-\sin^2\theta)
=
-2\cos\theta.
\end{align}
Thus \(R^2 n_z=-2n_z\).

\paragraph{(ii) \(n_x=\sin\theta\cos\phi\).}
Using
\(
\partial_\theta n_x=\cos\theta\cos\phi
\),
\(
\partial_\phi^2 n_x=-\sin\theta\cos\phi
\),
we find
\begin{align}
R^2 n_x
&=
\frac{1}{\sin\theta}\partial_\theta
(\sin\theta\cos\theta\cos\phi)
-
\frac{1}{\sin\theta}\cos\phi \nonumber\\
&=
-2\sin\theta\cos\phi
=
-2n_x.
\end{align}

\paragraph{(iii) \(n_y=\sin\theta\sin\phi\).}
By the same calculation,
\begin{equation}
R^2 n_y=-2n_y.
\end{equation}

Combining all components, we obtain
\begin{equation}
R^2\hat{\bm n}=-2\hat{\bm n},
\label{eq:R2n_final_app}
\end{equation}
which is the identity used in the main text.

\section{MSD calculation in dense regime $(\phi>\phi_c)$}\label{app:C}
\begin{align}
\Delta(t)
&= \left\langle \abs{\bm r(t)-\bm r(0)}^2 \right\rangle
= \langle r^2(t)\rangle
\nonumber\\
&=
\int_{0}^{t}\!\!\int_{0}^{t}
\left\langle \bm v(t_1)\cdot \bm v(t_2)\right\rangle\,dt_1\,dt_2 .
\label{eq1}
\end{align}

\noindent
where $\langle \bm v(t_1)\cdot \bm v(t_2)\rangle$ is the velocity autocorrelation function in Eq.~\ref{eq1}.

\medskip
\noindent
From the Langevin equation for position; deriving velocity:
\begin{align}
\dot{\bm r}(t) &= v_\rho(t)\,\hat{\bm n}(t) + \sqrt{2D}\,\bm\xi_\rho(t), \nonumber\\
\bm v(t) &= v_\rho(t)\,\hat{\bm n}(t) + \sqrt{2D}\,\bm\xi_\rho(t).
\label{eq2}
\end{align}

\medskip
\noindent
Splitting velocity autocorrelation into speed autocorrelation and orientation autocorrelation using Eq.~\ref{eq2}:
\begin{align}
\left\langle \bm v(t_1)\cdot \bm v(t_2)\right\rangle
&=
\left\langle v_\rho(t_1)\,v_\rho(t_2)\right\rangle
\left\langle \hat{\bm n}_1\cdot \hat{\bm n}_2\right\rangle
+ 6D\,\delta(t_1-t_2).
\label{eq3}
\end{align}

\noindent
In Eq.~\ref{eq3}: (i) $D$ is the translational diffusion constant; (ii) $\langle \hat{\bm n}_1\cdot \hat{\bm n}_2\rangle$ is the orientation correlation.

\medskip
\noindent
For $t_2>t_1$,
\begin{align}
\left\langle \hat{\bm n}_1\cdot \hat{\bm n}_2\right\rangle_{t_2>t_1}
&=
e^{-2D_R(t_2-t_1)}\,\frac{1}{3}\Big\{1+2\cos\big(\omega_0(t_2-t_1)\big)\Big\}.
\label{eq_orientcorr}
\end{align}

\noindent
(averaged over all possible initial orientations $\hat{\bm n}_0$ in 3D)

\medskip
\noindent
\textbf{Note:} Calculation of $\langle \hat{\bm n}_1\cdot \hat{\bm n}_2\rangle$ is separate and is
done using the Langevin equation and corresponding Fokker--Planck equation for orientation (in the manuscript):
\begin{align}
\partial_t \hat{\bm n} &=
\hat{\bm n}\times \omega_0 \hat{\bm z}
\;+\; \sqrt{2D_R}\,\bm\eta(t)\times \hat{\bm n},
\nonumber\\
\partial_t P &=
D_R\,\mathcal{R}^{2}P \;-\; \omega_0\,\mathcal{R}_{z}P.
\nonumber
\end{align}

\medskip
\noindent
(iii) $\langle v_\rho(t_1)\,v_\rho(t_2)\rangle$ $\rightarrow$ speed autocorrelation (as in Eq.\ref{eq:speed_dense_full})

\medskip
\noindent
Thus, we calculate the speed autocorrelation for the particle in the dense system in Eq.~\ref{eq3}.
We take the functional form of speed:
\begin{align}
v_\rho(t) = v_0\Big(1-\lambda_1\,\rho(t)+\lambda_2\,\rho^2(t)\Big),
\qquad \lambda_1,\lambda_2>0.
\label{eq4}
\end{align}

\noindent
Moment $\langle v_\rho(t_1)\,v_\rho(t_2)\rangle$ yields total \fbox{9} terms due to the quadratic form in Eq.~\ref{eq4}:
\begin{align}
\langle v_\rho(t_1)v_\rho(t_2)\rangle
= v_0^2\left(
1 - \underbrace{\cdots}_{\text{9 terms}}
\right).
\nonumber
\end{align}

\medskip
\noindent
In Eq.~\ref{eq4}, $\rho(t)$ is a stochastic variable with properties:
\begin{align}
\rho(t) &= \langle \rho(t)\rangle + \delta\rho(t), \nonumber\\[1mm]
\text{(i)}\quad
\langle \rho(t)\rangle
&= \rho_{0}\exp(-t/\tau)+\rho_{\infty}\Big[1-\exp(-t/\tau)\Big],
\label{eq5}
\end{align}
where $\rho_0$ is the initial local density at $t=0$ and $\rho_\infty$ is the constant local density achieved in the non-equilibrium steady state.

\begin{align}
\text{(ii)}\quad
C_{\rho}(t_1,t_2)
&=\left\langle \delta\rho(t_1)\,\delta\rho(t_2)\right\rangle
=\sigma_{\rho}^{2}\exp\!\left(-\frac{|t_1-t_2|}{\tau}\right).
\label{eq6}
\end{align}

\medskip
\noindent
\textbf{Note:} $\rho(t)$ is modeled similar to a nonstationary OU process with exponentially correlated
fluctuations to capture transient dynamics and steady-state fluctuations observed in simulations.

\medskip
\noindent
In the speed autocorrelation function expression's 9 terms, we:

\medskip
\noindent
(i) Neglect all three-point correlations, as they vanish according to Wick's theorem.
For zero-mean Gaussian fluctuations $\delta\rho_i$:
\begin{align}
\left\langle \delta\rho_1\,\delta\rho_2\cdots\delta\rho_n\right\rangle
&=
\sum\nolimits' \;\prod \left\langle \delta\rho_i\,\delta\rho_j\right\rangle,
 \text{ if $n$ is even},
\nonumber\\
\left\langle \delta\rho_1\,\delta\rho_2\cdots\delta\rho_n\right\rangle
&=0,\qquad \text{if $n$ is odd}.
\nonumber
\end{align}

\noindent
(ii) We neglect higher-order terms in density and retain only first-order terms in the mean $\langle\rho(t)\rangle$
and correlation $C_{\rho}(t_1,t_2)=\langle \delta\rho(t_1)\delta\rho(t_2)\rangle$,
as they contribute strongly to the MSD.
This truncation retains contributions up to second order in density fluctuations and neglects higher-order cumulants, consistent with a lowest-order closure approximation.\\
\medskip
\noindent
(Starting the expansion exactly as written at the bottom of page 2:)

\onecolumngrid

\begin{align}
\langle v_\rho(t_1)\,v_\rho(t_2)\rangle
&= v_0^{2}\,
\Big[\big(1-\lambda_1\langle\rho_1\rangle-\lambda_1\delta\rho_1
+\lambda_2(\langle\rho_1\rangle^{2}+\delta\rho_1^{2}+2\langle\rho_1\rangle\delta\rho_1)\big)\Big]
\nonumber\\
&\quad\times
\Big[\big(1-\lambda_1\langle\rho_2\rangle-\lambda_1\delta\rho_2
+\lambda_2(\langle\rho_2\rangle^{2}+\delta\rho_2^{2}+2\langle\rho_2\rangle\delta\rho_2)\big)\Big].
\nonumber
\end{align}


\noindent
\underline{Speed autocorrelation function}
\begin{align}
\langle v_\rho(t_1)\,v_\rho(t_2)\rangle
&\simeq
v_0^{2}
\Big[
1
-\lambda_1\{1
+2\lambda_2\,C_{\rho}(t_1,t_2)\}(\langle\rho_1\rangle+\langle\rho_2\rangle)
+\lambda_1^{2}\,C_{\rho}(t_1,t_2)
\Big].
\label{eq7}
\end{align}

\medskip
\noindent
From speed autocorrelation in Eq.~\ref{eq7}, we calculate velocity autocorrelation in Eq.~\ref{eq3}
and use it to calculate MSD in Eq.~\ref{eq1}:
\begin{align}
\langle r^{2}\rangle
&=
\int_{0}^{t}\!\!\int_{0}^{t}
\langle \bm v(t_1)\cdot \bm v(t_2)\rangle \, dt_1 dt_2
\nonumber\\[1mm]
&=
\int_{0}^{t}\!\!\int_{0}^{t}
\langle v_\rho(t_1)v_\rho(t_2)\rangle
\langle \hat{\bm n}_1\cdot\hat{\bm n}_2\rangle
\,dt_1dt_2
\;+\;
6D\int_{0}^{t}\!\!\int_{0}^{t}
\delta(t_1-t_2)\,dt_1dt_2
\nonumber\\[1mm]
&=
6Dt
+
v_0^{2}
\int_{0}^{t}\!\!\int_{0}^{t}
\Big[
1
-\lambda_1\{1+
2\lambda_2\,C_{\rho}(t_1,t_2)\}(\langle\rho_1\rangle+\langle\rho_2\rangle)
+\lambda_1^{2}\,C_{\rho}(t_1,t_2)
\Big]
\langle \hat{\bm n}_1\cdot\hat{\bm n}_2\rangle
\,dt_1dt_2 
\nonumber\\[1mm]
&=
6Dt+\Delta_{0}(t)+\Delta_{T}(t)+\Delta_{S}(t).
\nonumber
\end{align}

\medskip
\noindent
where,
\begin{align}
\Delta_{0}(t) &= \text{Mean squared displacement of a chiral ABP in 3D without any interactions}, \nonumber\\
\Delta_{T}(t) &= \text{MSD contribution due to kinetic growth of clusters}, \nonumber\\
\Delta_{S}(t) &= \text{MSD contribution due to steady-state fluctuations in clusters}. \nonumber
\end{align}

\medskip
\noindent
\begin{align}
\text{(I)}\quad
\Delta_{0}(t)
&=
v_0^{2}
\int_{0}^{t}\!\!\int_{0}^{t}
\langle \hat{\bm n}_1\cdot\hat{\bm n}_2\rangle
\,dt_1dt_2 ,
\nonumber\\[2mm]
\text{(II)}\quad
\Delta_{T}(t)
&=
-\lambda_1 v_0^{2}
\int_{0}^{t}\!\!\int_{0}^{t}
\{1+2\lambda_2C_{\rho}\}\{\langle\rho_1\rangle+\langle\rho_2\rangle\}
\langle \hat{\bm n}_1\cdot\hat{\bm n}_2\rangle
\,dt_1dt_2,
\nonumber\\[2mm]
\text{(III)}\quad
\Delta_{S}(t)
&=
\lambda_1^{2} v_0^{2}
\int_{0}^{t}\!\!\int_{0}^{t}
C_{\rho}(t_1,t_2)
\langle \hat{\bm n}_1\cdot\hat{\bm n}_2\rangle
\,dt_1dt_2 .
\nonumber
\end{align}


\medskip
\noindent
\textbf{(I) Calculation of $\Delta_0(t)$:}
\begin{align}
\Delta_0(t)
&=
v_0^{2}\int_{0}^{t}\!\!\int_{0}^{t}\,
\langle \hat{\bm n}_1\cdot \hat{\bm n}_2\rangle\,dt_1\,dt_2 ,
\nonumber
\end{align}
where
\begin{align}
\langle \hat{\bm n}_1\cdot \hat{\bm n}_2\rangle
=
\frac{1}{3}\exp\!\left[-2D_R|t_2-t_1|\right]
\left\{1+2\cos\!\big(\omega_0|t_2-t_1|\big)\right\}.
\nonumber
\end{align}

\medskip
\noindent
Now split the double integral into two parts:
\begin{align}
\Delta_0(t)
&=
v_0^2\left[
\underbrace{
\int_{0}^{t}dt_2\int_{0}^{t_2}dt_1\,
\langle \hat{\bm n}_1\cdot \hat{\bm n}_2\rangle_{t_1<t_2}
}_{\text{(A)}}
+
\underbrace{
\int_{0}^{t}dt_2\int_{t_2}^{t}dt_1\,
\langle \hat{\bm n}_1\cdot \hat{\bm n}_2\rangle_{t_1<t_2}
}_{\text{(B)}}
\right].
\nonumber
\end{align}

\noindent
Term (A) and Term (B) are equal since $\langle \hat{\bm n}_1\cdot \hat{\bm n}_2\rangle$ is symmetric about $t_1=t_2$.

\begin{align}
\therefore\quad
\Delta_0(t)
&=
2v_0^2\int_{0}^{t}dt_2\int_{0}^{t_2}dt_1\,
\langle \hat{\bm n}_1\cdot \hat{\bm n}_2\rangle_{t_1<t_2}
\nonumber\\
&=
2v_0^2\int_{0}^{t}dt_2\int_{0}^{t_2}dt_1\,
\frac{1}{3}\exp[-2D_R(t_2-t_1)]
\left\{1+2\cos\!\omega_0(t_2-t_1)\right\}
\nonumber\\
&=
\frac{2v_0^2}{3}\int_{0}^{t}dt_2\,e^{-2D_R t_2}
\int_{0}^{t_2}dt_1\,e^{2D_R t_1}
\left\{1+2\cos\!\omega_0(t_2-t_1)\right\}
\nonumber\\
&=
\frac{2v_0^2}{3}\int_{0}^{t}dt_2\,e^{-2D_R t_2}
\int_{0}^{t_2}dt_1\,e^{2D_R t_1}
+\frac{4v_0^2}{3}\int_{0}^{t}dt_2
\int_{0}^{t_2}dt_1\,e^{-2D_R(t_2- t_1)}\cos\!\omega_0(t_2-t_1)
\nonumber\\
&=
\frac{2v_0^2}{3}\int_{0}^{t}dt_2\,e^{-2D_R t_2}
\left[
\frac{e^{2D_R t_2}-1}{2D_R}
\right]
-\frac{4v_0^2}{3}\int_{0}^{t}dt_2\int_{t_2}^{0}du\,e^{-2D_Ru}\cos\!\omega_0u.
\nonumber
\end{align}

\medskip
\noindent
$u=(t_2-t_1)$.


\begin{align}
\Delta_0(t)
&=
\frac{v_0^{2}}{3D_R}\int_{0}^{t}dt_2\Big[1-e^{-2D_R t_2}\Big]
-\frac{4v_0^{2}}{3}\int_{0}^{t}dt_2\int_{t_2}^{0}du\;
e^{-2D_R u}\cos(\omega_0 u).
\nonumber
\end{align}

\noindent
For the cosine part,
\begin{align}
\int_{t_2}^{0}du\,e^{-2D_R u}\cos(\omega_0 u)
&=
\left[
\frac{e^{-2D_R u}}{4D_R^2+\omega_0^2}
\Big\{\omega_0\sin(\omega_0 u)-2D_R\cos(\omega_0 u)\Big\}
\right]_{u=t_2}^{u=0}
\nonumber\\
&=
\frac{1}{4D_R^2+\omega_0^2}
\Big[
(-2D_R)-e^{-2D_R t_2}\big\{\omega_0\sin(\omega_0 t_2)-2D_R\cos(\omega_0 t_2)\big\}
\Big].
\nonumber
\end{align}

\noindent
Substituting back,
\begin{align}
\Delta_0(t)
&=
\frac{v_0^{2}}{3D_R}\int_{0}^{t}dt_2\Big[1-e^{-2D_R t_2}\Big]
\nonumber\\
&\quad
+\frac{4v_0^{2}}{3}\frac{1}{4D_R^2+\omega_0^2}
\int_{0}^{t}dt_2
\Big[
(-2D_R)
-e^{-2D_R t_2}\big\{\omega_0\sin(\omega_0 t_2)-2D_R\cos(\omega_0 t_2)\big\}
\Big]
\nonumber\\
&=
\frac{v_0^{2}}{6D_R^{2}}
\Big[e^{-2D_R t}+2D_R t-1\Big]
\nonumber\\
&\quad
+\frac{4v_0^{2}}{3}\frac{1}{(4D_R^2+\omega_0^2)^{2}}
\Big[
2D_R(4D_R^2+\omega_0^2)t
+e^{-2D_R t}\big\{(4D_R^2-\omega_0^2)\cos(\omega_0 t)-4\omega_0 D_R\sin(\omega_0 t)\big\}
+\big(\omega_0^2-4D_R^2\big)
\Big].
\label{eq8}
\end{align}


\noindent
For convenience, define the integral of the orientation autocorrelation
\begin{align}
I(t)\equiv \int_{0}^{t}\!\!\int_{0}^{t}\langle \hat{\bm n}_1\cdot \hat{\bm n}_2\rangle\,dt_1dt_2,
\nonumber
\end{align}

whose value is given in Eq.\ref{eq8}. Note~$\rightarrow$~ $I(t)=\Delta_0(t)/v_0^2$ and $\Delta_0(t)$ is known from Eq.\ref{eq8}. Value of $I(t)$ will be used in calculation of $\Delta_S(t)$ and $\Delta_T(t)$.

\medskip
\noindent
\textbf{(III) Calculation of $\Delta_S(t)$:}
\begin{align}
\Delta_S(t)
&=
\lambda_1^2 v_0^2\int_{0}^{t}\!\!\int_{0}^{t}
C_{\rho}(t_1,t_2)\,\langle \hat{\bm n}_1\cdot \hat{\bm n}_2\rangle\,dt_1dt_2 .
\nonumber
\end{align}

\noindent
From Eq.~\ref{eq6},
$C_{\rho}(t_1,t_2)=\sigma_{\rho}^{2}\exp\!\left[-\frac{|t_1-t_2|}{\tau}\right]$.
From Eq.~\ref{eq_orientcorr},
$\langle \hat{\bm n}_1\cdot \hat{\bm n}_2\rangle=
\frac{1}{3}\exp[-2D_R|t_1-t_2|]\{1+2\cos(\omega_0|t_1-t_2|)\}$.

\medskip
\noindent
Thus,
\begin{align}
\Delta_S(t)
&=
\lambda_1^2 v_0^2 \sigma_{\rho}^{2}
\int_{0}^{t}\!\!\int_{0}^{t}
\frac{1}{3}\exp\!\left[-\left(2D_R+\frac{1}{\tau}\right)|t_1-t_2|\right]
\Big\{1+2\cos\big(\omega_0|t_1-t_2|\big)\Big\}\,dt_1dt_2
\nonumber\\
&\equiv
\lambda_1^2 v_0^2 \sigma_{\rho}^{2}\, I_{\beta}(t),
\label{eq9}
\end{align}
where $I_\beta(t)$ is $I(t)=\int_{0}^{t}\!\!\int_{0}^{t}\langle \hat{\bm n}_1\cdot \hat{\bm n}_2\rangle\,dt_1dt_2$; but with transformation/substitution of $2D_R\rightarrow2D_R+1/\tau = \beta$.

\medskip
\noindent
Hence,
\begin{align}
\Delta_S(t)
&=
\lambda_1^2\sigma_{\rho}^2\frac{2v_0^2}{3\beta^2}\Big[\beta t+e^{-\beta t}-1\Big]
\nonumber\\
&\quad
+\lambda_1^2\sigma_{\rho}^2\frac{4v_0^2}{3(\beta^2+\omega_0^2)^{2}}
\Big[
\omega_0^2-\beta^2
+\beta(\beta^2+\omega_0^2)t
+e^{-\beta t}\big((\beta^2-\omega_0^2)\cos\omega_0 t-2\omega_0\beta \sin\omega_0 t\big)
\Big].
\label{eq10}
\end{align}


\medskip
\noindent
\textbf{(II) Calculation of $\Delta_T(t)$:}
\begin{align}
\Delta_T(t)
&=
-\lambda_1 v_0^{2}\int_{0}^{t}\!\!\int_{0}^{t}
\Big\{1 + 2\lambda_2\,C_{\rho}\Big\}\,
\Big(\langle\rho_1\rangle+\langle\rho_2\rangle\Big)\,
\langle \hat{\bm n}_1\!\cdot\!\hat{\bm n}_2\rangle\;dt_1dt_2.
\nonumber
\end{align}

\noindent
Using Eq.~\ref{eq6}: $C_{\rho}(t_1,t_2)=\sigma_{\rho}^{2}\exp\!\left[-\frac{|t_1-t_2|}{\tau}\right]$,
and Eq.~\ref{eq5}: $\langle\rho(t)\rangle=\rho_{\infty}-(\rho_{\infty}-\rho_{0})e^{-t/\tau}$, we obtain:
\begin{align}
\Delta_T(t)
&=
-2\lambda_1\rho_{\infty}v_0^{2}\int_{0}^{t}\!\!\int_{0}^{t}
\langle \hat{\bm n}_1\!\cdot\!\hat{\bm n}_2\rangle\,dt_1dt_2
\nonumber\\
&\quad
+\lambda_1(\rho_{\infty}-\rho_{0})v_0^{2}\int_{0}^{t}\!\!\int_{0}^{t}
\Big(e^{-t_1/\tau}+e^{-t_2/\tau}\Big)\,
\langle \hat{\bm n}_1\!\cdot\!\hat{\bm n}_2\rangle\,dt_1dt_2
\nonumber\\
&\quad
-4\lambda_1\lambda_2\rho_{\infty}\sigma_{\rho}^{2}v_0^{2}\int_{0}^{t}\!\!\int_{0}^{t}
e^{-|t_1-t_2|/\tau}\,
\langle \hat{\bm n}_1\!\cdot\!\hat{\bm n}_2\rangle\,dt_1dt_2
\nonumber\\
&\quad
+2\lambda_1\lambda_2\sigma_{\rho}^{2}(\rho_{\infty}-\rho_{0})v_0^{2}\int_{0}^{t}\!\!\int_{0}^{t}
e^{-|t_1-t_2|/\tau}\,
\langle \hat{\bm n}_1\!\cdot\!\hat{\bm n}_2\rangle\,
\Big(e^{-t_1/\tau}+e^{-t_2/\tau}\Big)\,dt_1dt_2 .
\label{eq11}
\end{align}

\medskip
\noindent
\textbf{Note:} If
\begin{align}
\int_{0}^{t}\!\!\int_{0}^{t}\langle \hat{\bm n}_1\!\cdot\!\hat{\bm n}_2\rangle\,dt_1dt_2
= I(t),
\nonumber
\end{align}
then define
\begin{align}
\int_{0}^{t}\!\!\int_{0}^{t}
\langle \hat{\bm n}_1\!\cdot\!\hat{\bm n}_2\rangle\,e^{-t_1/\tau}
\,dt_1dt_2
= I_{\mathrm{filt}(1/\tau)}(t),
\nonumber
\end{align}
where
\begin{align}
I_{\mathrm{filt}(1/\tau)}(t)
= I(t) - J^{0}(t).
\label{eq12}
\end{align}

\begin{align}
J^{0}(t)
&=
\frac{1}{6D_R^{2}}
\Bigg[
2D_R\tau\Big(1-e^{-t/\tau}\Big)
+\frac{e^{-2D_R t}-e^{-t/\tau}}{(1/\tau)-2D_R}
\Bigg]
\nonumber\\
&\quad
+\frac{4}{3}\,\frac{1}{(4D_R^{2}+\omega_0^{2})}
\Bigg[
A\Big(1-e^{-t/\tau}\Big)
+\frac{B\,(e^{-2D_R t}-e^{-t/\tau})}{(1/\tau^{2})+\omega_0^2}
\Bigg].
\nonumber
\end{align}


\noindent
where the constants in $J^{0}(t)$ are
\begin{align}
A
&=
\frac{\omega_0^{2}-4D_R^{2}}{1/\tau}
+
\frac{2D_R(4D_R^{2}+\omega_0^{2})}{1/\tau^{2}+\omega_0^{2}},
\nonumber\\
B
&=
\frac{(4D_R^{2}-\omega_0^{2})}{\tau}
-4D_R\omega_0 .
\nonumber
\end{align}

\medskip
\noindent
Substituting Eq.~\ref{eq11} into the definition of $\Delta_T(t)$,
\begin{align}
\Delta_T(t)
&=
-2\lambda_1\rho_{\infty}v_0^{2} I(t)
+2\lambda_1(\rho_{\infty}-\rho_{0})v_0^{2}
I_{\mathrm{filt}(1/\tau)}(t)
\nonumber\\
&\quad
-4\lambda_1\lambda_2\rho_{\infty}\sigma_{\rho}^{2}v_0^{2} I_{\beta}(t)
+4\lambda_1\lambda_2(\rho_{\infty}-\rho_{0})\sigma_{\rho}^{2}v_0^{2}
I_{\beta,\mathrm{filt}(1/\tau)}(t).
\label{eq13}
\end{align}

\noindent
Here $J_{\beta}^{0}(t)$ (and hence $I_{\beta,\mathrm{filt}(1/\tau)}$) is defined analogously to $J^{0}(t)$,
with the replacement $2D_R \rightarrow \beta$.

\medskip
\noindent
Finally,
\begin{align}
\langle r^{2}\rangle
&=
6Dt+\Delta_0(t)+\Delta_T(t)+\Delta_S(t),
\nonumber
\end{align}
using Eqs.~\ref{eq8}, \ref{eq10}, and \ref{eq13} we will get the full expression of $<r^2(t)>$ as given in Eq.\ref{msd_dense} in the main manuscript.

\twocolumngrid

\end{document}